\documentclass[pdftex,twocolumn,epjc3]{svjour3}          

\RequirePackage[T1]{fontenc}

\smartqed  

\RequirePackage{graphicx}
\usepackage{latexsym}
\usepackage{multirow}
\usepackage{amsmath,amssymb,amsfonts,latexsym,cancel}
\usepackage{subcaption}
\RequirePackage{mathptmx}      
\RequirePackage{flushend}
\RequirePackage[numbers,sort&compress]{natbib}
\RequirePackage[colorlinks,citecolor=blue,urlcolor=blue,linkcolor=blue]{hyperref}

\journalname{Eur. Phys. J. C}

\begin{document}

\title{Phase transition and stiffer core fluid in neutron stars: Effects on stellar configurations, dynamical stability, and tidal deformability}

\author{Jos\'e D. V. Arba\~nil\thanksref{e1,addr1,addr2}
        \and
        Lucas S. Rodrigues\thanksref{addr3}
        \and
        César H. Lenzi\thanksref{addr3}  
        }
\thankstext{e1}{e-mail: jose.arbanil@upn.pe}

\institute{Departamento de Ciencias, Universidad Privada del Norte, Avenida el Sol 461 San Juan de Lurigancho, 15434 Lima, Peru \label{addr1} \and
Facultad de Ciencias F\'isicas, Universidad Nacional Mayor de San Marcos, Avenida Venezuela s/n Cercado de Lima, 15081 Lima,  Peru \label{addr2}
\and
Departamento de F\'isica, Instituto Tecnol\'ogico de Aeron\'autica, S\~ao Jos\'e dos Campos, SP, 12228-900, Brazil
\label{addr3}
}

\date{Received: date / Accepted: date}

\maketitle

\begin{abstract}
In this work, we investigate the influence of the phase transition and a stiffer fluid in neutron stars' cores on the static equilibrium configuration, dynamical stability, and tidal deformability. For this aim, it is taken into account that the fluid in the core and the envelope follow the relativistic polytropic equation of state. We find that the phase transition and a stiffer fluid in the core will reflect in the total mass, radius, speed of sound, core radius, radial stability with a slow and rapid conversion at the interface, and tidal deformability. We also investigate the dimensionless tidal deformability $\Lambda_1$ and $\Lambda_2$ for a binary neutron stars system with chirp mass equal to GW$170817$. Finally, we contrast our results with observational data to show the role that phase transition and a stiffer core fluid could play in the study of neutron stars. 
\end{abstract}
\maketitle


\section{Introduction}\label{sec-introd}

The direct multimessenger detection from binary black holes merger carried out by the LIGO-Virgo scientific network \cite{abbott2016,abbott2016_2,abbott2017,abbott2017_1,abbott2017_2} has marked the starting of the era of Gravitational Waves (GWs) astronomy. The detection of GWs has opened a new window to explore the cosmos and supplied some astrophysics and fundamental physics implications (check, e.g., \cite{abbott2017_3,abbott2016_3,yunes2016}). Another important event of GWs comes from a merger of a pair of neutron stars (NSs) \cite{abbott2017_NS}, known as event GW$170817$, which was also reported by the LIGO-Virgo scientific network. This new signal opened the GWs multi-messenger astronomy, being the first detection with electromagnetic counterpart \cite{abbott2017_4}, and providing a set of valuable information about the properties of NSs and their equation of state (EOS). After the first detection of GWs from the NSs binary system, many important efforts have been realized to constrain, e.g., their radii and EOS \cite{malik_2018,most_2018,annala_2018,paschalidis_2018,lugones2019}. Additional constraints are feasible because of the implication of tidal deformation \cite{hinderer_2008,damour_2009,hinderer_2010}.

It is known that the matter density that makes up NSs reach densities up to a few times the nuclear saturation density, however, until these days, detailed information about the characteristics and nature of their deep interiors is still lacking. Future multi-messenger signatures hold the promise of identifying the specific internal aspect of NSs. Theoretically, asteroseismology is widely employed to analyze the internal structure of compact stars -the name used for white dwarfs, neutron stars, hybrid stars, or strange quark stars- to investigate the thermodynamic properties inside these objects. Through this diagnostic technique, analyzing the frequency modes can obtain a solid way to learn more about the physics inside compact stars. For example, if inside these stars a single component fluid is present \cite{chanmugam1977,Alcock_1986,vath_chanmugam1992,bombaci1996,gondek1997,gondek1999,kokkotas2001,lugones2010,doneva2012,arbanil_malheiro2015,arbanil_malheiro_2016,flores2017,sagun_2020,annala_2022} or the existence of a phase transition between layers with different mechanical properties \cite{pereira_flores2018,tonetto_2020,miniutti_2003,mishustin_2003,sahu2002,gupta2001,flores_lenzi2012,brillante_2014,clemente_2020,jimenez_2021,parisi_2021,sun_2021,mariani_lugones2019,ivanytskyi_2022}. 

In literature, unlike the study of one-phase static compact stars, two-phase stars and the impact of the phase transition on the properties of these stars were not widely investigated. There are studies analyzing how density jumps affect the static stellar equilibrium configuration and radial frequency of oscillations \cite{sahu2002,gupta2001,flores_lenzi2012,brillante_2014,clemente_2020,jimenez_2021,sun_2021,parisi_2021,mariani_lugones2019,pereira_flores2018,tonetto_2020,mishustin_2003}, as well as the possibility of arising of the so-called gravitational pulsation mode ($g$-mode) \cite{miniutti_2003,mishustin_2003,tonetto_2020}. 
 
As regards the radial perturbations of compact stars with a sharp interface, the set of equations must be solved by taking into account the additional boundary conditions at the phase-splitting interface. Around this point, there are two types of physical behavior due to radial perturbations: the slow and rapid phase conversion \cite{pereira_flores2018,tonetto_2020}. In the case of slow conversion, there is no change of matter over the pulsating interface. On the contrary, the rapid conversion case involves a flow of mass from one phase to the other, and vice-versa, through the moving phase boundary. In recent years, the impact of the phase transition on the radial oscillations of compact stars has been reported in different articles. In the case of slow transition, for example, the authors concentrate on investigating the effects of the core formation \cite{mishustin_2003}, a sharp phase transition \cite{clemente_2020,parisi_2021}, the mixed-phase \cite{sahu2002,gupta2001,flores_lenzi2012,sun_2021,jimenez_2021}, and electric charge \cite{brillante_2014}. On the other hand, among those reported considering both phase transitions in compact stars, we find: the ones that analyze the fast and slow conversion in the context of general relativity \cite{pereira_flores2018}, how these are affected in the face of a magnetic field \cite{mariani_lugones2019}, and their influences on non-radial oscillations \cite{tonetto_2020}. In the rapid phase transition case, unlike the slow phase transition, in a sequence of equilibrium configuration, the maximum mass peak marks the beginning of radial instability. Indeed, in slow transitions, after this turning point, it is possible to find additional stable equilibrium configurations. Therefore, in a sequence of equilibrium configurations with increasing central energy density, some stars with the same mass but different radii are obtained. These stars are known as twin stars.

In the aforementioned articles, different models of equations of state are studied from the perspective of observational deformability data from the event GW170917. Some of these works study this phenomenon against the possibility of the existence of phase transitions inside the compact star, some taking the aspect of an analysis of the stability of these stars, and calculating radial oscillations. However, these same articles often make these calculations assuming only slow transitions, which allow the appearance of stable regions, in the mass-radius diagram, after the maximum mass. In this work, we present a detailed study of the influence of the phase transition and a stiffer fluid in NSs core on the equilibrium configuration, radial stability, and tidal deformability. In this sense, we analyze how the radius, mass, speed of sound, core radius, radial frequency of oscillation, and tidal deformation change when a phase transition and stiffer fluid in NSs core are considered. In the analysis of the radial stability of NSs, we will focus on the slow and rapid phase conversions. We also contrast our results with observational data to see the role that phase transition and a stiffer core fluid could play in the study of NSs.

The present article is arranged as follows: Section \ref{sec-basicequations} presents the equilibrium equations and radial stability equations; moreover, this section is also devoted to presenting the junction conditions at the interface of the two-phase, which are required to investigate the slow and rapid phase conversions. Section \ref{eos_employed} presents the EOSs employed for NSs, as well as the numerical method used to solve the complete set of equations required to investigate the equilibrium and radial stability. In Section \ref{results} we show the numerical results for equilibrium configurations, radial stability, and tidal deformability of NSs with two-phase. Finally, we conclude in Section \ref{conclusion}. Throughout the paper, we work with geometric units, i.e., $c=1=G$, and the metric signature $+2$.

\section{General relativistic formulations}\label{sec-basicequations}

\subsection{Equilibrium equations}

We take into account that the unperturbed neutron star is made up of layers of effective perfect fluids, whose energy-momentum tensors can be expressed as
\begin{equation}\label{tem}
T_{\mu\nu}=\left(\rho+p\right)\,u_{\mu}\,u_{\nu}+p\,g_{\mu\nu},
\end{equation}
with $\rho$, $p$, and $u_{\mu}$ representing respectively the energy density, the fluid pressure, and the four-velocity.

To analyze the effect of phase transition in the dense matter on the equilibrium and radial stability of neutron stars, we set the space-time metric, in Schwarzschild coordinates, as
\begin{equation}\label{metric}
ds^2=-e^{\nu}dt^2+e^{\beta}dr^2+r^2d\theta^2+r^2\sin^2\theta\,d\phi^2.
\end{equation}
The potential metric functions $\nu=\nu(r)$ and $\beta=\beta(r)$ depend on the radial coordinate $r$ only.

For the energy-momentum tensor (Eq.~\eqref{tem}) and line element (Eq.~\eqref{metric}) adopted, with the potential metric $e^{-\beta}=\left(1-2m/r\right)$, we derive the set of stellar structure equations
\begin{eqnarray}
&&\frac{dm}{dr}=4\pi\rho r^2,\label{eq_masa}\\
&&\frac{dp}{dr}=-(p+\rho)\left(4\pi rp+\frac{m}{r^2}\right)e^{\beta},\label{tov2}\\
&&\frac{d\nu}{dr}=-\frac{2}{(p+\rho)}\frac{dp}{dr},\label{eq_nu}
\end{eqnarray}
where the parameter $m$ represents the mass inside the sphere of radius $r$. Eq.~\eqref{tov2} is known as the hydrostatic equilibrium equation for a spherically symmetric static astrophysical object, also called as Tolman-Oppenheimer-Volkoff equation \cite{tolman,oppievolkoff}. 

The stellar structure equations \eqref{eq_masa}-\eqref{eq_nu} are integrated from the center toward the star's surface. At the center $(r=0)$ the integration starts
\begin{equation}
m(0)=0,\;\;p(0)=p_c,\;\;\rho(0)=\rho_c,\;{\rm and}\;\nu(0)=\nu_c.
\end{equation}
The surface of the star $(r=R)$ is determined by
\begin{equation}\label{null_pressure}
p(R)=0.
\end{equation}
At this point, the interior solution connects smoothly with the exterior Schwarzschild vacuum solution. This indicates that at the star's surface the interior and exterior potential metrics are related through of the form:
\begin{equation}
e^{\nu(R)}=e^{-\beta(R)}=1-\frac{2M}{R},
\end{equation}
with $M$ being the total mass of the star.

\subsection{Radial oscillations equations}

The radial pulsation equation is obtained by Chandrasekhar \cite{chandrasekhar_rp} making perturbation in the fluid and space-time variables. The perturbed quantities are placed into Einstein's field equation and in the linearized form of the conservation of stress-energy tensor.

The solution of the radial oscillation equation provides information about the eigenfrequency of oscillations $\omega$. Intending to set this equation in a more appropriate form for numerical integration, we place it into two first-order equations for the variables $\Delta r/r$ and $\Delta p$ \cite{gondek1999}; with $\Delta r$ and $\Delta p$ representing respectively the relative radial displacement and Lagrangian perturbations of pressure. Thus, the system of equations is, $\xi = \Delta r/r$:
\begin{eqnarray}
&&\frac{d\xi}{dr}=\frac{\xi}{2}\frac{d\nu}{dr}-\frac{1}{r}\left(3\xi+\frac{\Delta p}{p\Gamma}\right),\label{ro1}\\
&& \ \ \ \ \ \ \ \nonumber \\
&&\frac{d\Delta p}{dr}= (p+\rho)\omega^2\xi re^{\beta-\nu}-4\xi\left(\frac{dp}{dr}\right)+\left(\frac{dp}{dr}\right)^2\frac{\xi r}{p+\rho}\nonumber\\
&&-8\pi p\,(p+\rho)\xi r e^{\beta} -\left(\frac{1}{2}\frac{d\nu}{dr}+4\pi re^\beta(p+\rho)\right)\Delta p,\label{ro3} 
\end{eqnarray}
where $\Gamma=\left(1+\frac{\rho}{p}\right)\frac{dp}{d\rho}$. The variables $\xi$ and $\Delta p$ have a time dependence of the form $e^{i\omega t}$, with $\omega$ being the eigenfrequency.

To solve the differential equations \eqref{ro1} and \eqref{ro3}, boundary conditions in the center and on the star's surface are required. Moreover, to find regular solutions in the center of the star, the second term of the right-hand side of Eq.~\eqref{ro1} must vanish in $r\rightarrow 0$. In this way, it is considered
\begin{equation}
\left(\Delta p\right)_{\rm center}=-3\left(\xi\Gamma p\right)_{\rm center}.
\end{equation}
At this point, for normalized eigenfunctions, we regard $\xi(r=0)=1$. On the other hand, as established above, the surface of the star is determined when $p(R)=0$. It implies
\begin{equation}\label{delta_p}
\left(\Delta p\right)_{\rm surface}=0.
\end{equation}

\subsection{Tidal deformability}

Tidal effects are very common in the context of NSs binary systems. In fact, the gravitational field generated by one star in a binary system can result in deformation in its companion. The parameter of tidal deformability is the measure of the deformation in compact stars due to an external field. From a mathematical point of view, this parameter can be expressed in terms of the fraction, 
\begin{equation}
\lambda_1=-\frac{Q_{ij}}{\epsilon_{ij}}, 
\end{equation}
where $Q_{ij}$ is the quadrupole moment perturbed by an external tidal field $\epsilon_{ij}$ \cite{hinderer_2008,damour_2009,hinderer_2010}. The tidal deformability parameter $\lambda_1$ is connected with the Love number $k_2$ through the relation $k_2=\frac{3}{2}\lambda_1R^{-5}$. Moreover, the dimensionless tidal deformability $\Lambda$ can be written in terms of the Love number $k_2$ as 
\begin{equation}
\Lambda = \frac{2}{3}\frac{k_2}{C^5},
\end{equation}
with $C=M/R$ being the compactness parameter. $k_2$ can be expressed in terms of parameter $y_R \equiv y(r=R)$ as follows
\begin{equation}
\begin{aligned}
&{k_2}=\frac{8C^5}{5}(1-2C)^2[2+C(y_R-1)-y_R]\\
&\times\{2C [6-3y_R+3C(5y_R-8)]+4C^3[13-11y_R\\
&+C(3y_R-2)+2C^2(1+y_R)]+3(1-2C^2)\\
& \times [2-y_R+2(y_R-1)]\ln(1-2C)\}^{-1}. 
\end{aligned}
\end{equation}
The parameter $y(r)$ is calculated along the whole of the star -from the center to the surface of the star- integrating the equation
\begin{equation}\label{fff}
r \frac{dy}{dr} + y^2 + y F + r^2 Q = 0,
\end{equation}
together with the set of equations \eqref{eq_masa}-\eqref{eq_nu}, considering at the center of the star $y(r=0) = 2$. The functions $F=F(r)$ and $Q=Q(r)$ are represented by the relations:
\begin{eqnarray}
&&F = (\rho-p)\frac{r-4\pi r^3}{r-2m},\label{qqq}\\ 
&&Q = 4\pi e^{\beta} \left( 5 \rho + 9 p + \frac{\rho+p}{dp/d\rho}\right) - \frac{6e^{\beta}}{r^2}-\left( \frac{d\nu}{dr} \right)^2.\label{qqq1}
\end{eqnarray}

\subsection{Junction conditions at the interface}\label{jc_interface}

In the last years, compact stars with two different phases have been considered a real possibility. However, there are still a bunch of open questions such as, for instance, the density at which a hadron-quark phase transition occurs and some discussions about the kind of phase transition depending on the surface tension between the phases \cite{pereira_flores2018,tonetto_2020,Ilda1998,mariani_lugones2019,T-endo2011}. In our case, a first-order phase transition is considered, which results in the presence of a finite energy density discontinuity. Moreover, some approaches are regarded to investigate the radial stability and deformability of stars.

\subsubsection{Radial oscillations}

The phase transitions can be classified as slow or rapid depending of the time scale of the reaction of the matter in the neighborhood of the hadron-quark interface \cite{haensel1989,pereira_flores2018}.
A scenario of slow phase transition appears when the rate of reaction transforming one phase into another is much greater than those of the radial perturbations. In such circumstances, there is no flow of matter across the surface splitting the two phases. Such a condition implies that $\xi$ must always be continuous at the interface across the interface, i.e., 
\begin{equation}\label{slow_condition_1}
\xi_{\rm inn}=\xi_{\rm out}.
\end{equation}
Moreover, this also leads to a continuity pressure at the interface of the two phases. At this point, it assures that 
\begin{equation}\label{slow_condition_2}
\left(\Delta p\right)_{\rm inn}=\left(\Delta p\right)_{\rm out}.
\end{equation}

In the case of a rapid phase transition, the rate of reaction transforming one phase into another is much lower than those of the radial perturbations. In this scenario, a change of mass flow through the interface occurs. The diminution of mass on one side should be equal to the increase of mass on the other side. This condition, together with the demand for the continuity of pressure, lead to
\begin{equation}\label{rapid_condition_1}
\left(\xi-\frac{\Delta p}{rp'}\right)_{\rm inn}=\left(\xi-\frac{\Delta p}{rp'}\right)_{\rm out},
\end{equation}
where the prime stands the operation with respect to the radial derivative, and
\begin{equation}\label{rapid_condition_2}
\left(\Delta p\right)_{\rm inn}=\left(\Delta p\right)_{\rm out}.
\end{equation}

\subsubsection{Tidal deformability}

At the interface of the two layers, we can clearly note that exists a singularity in Eq. (\ref{qqq1}) due to the speed of sound ($dp/d\rho$) since, at this point, we have the same value of the fluid pressure for two different energy densities. In Ref. \cite{damour_2009}, the authors discussed this problem in the context of the surface vacuum discontinuity for incompressible stars. After in Refs \cite{postnikov_2010,takatsy_2020,zhang_2002}, this approach was extended for the case of first-order transitions inside hybrid-stars, where authors concluded that at this point the function $y(r)$ must follow the next condition:
\begin{equation}\label{defor_junc}
    y(r_{tr}+\epsilon) = y(r_{tr}-\epsilon) - \frac{4\pi r_{tr}^3\left[\rho(r_{tr}+\epsilon) - \rho(r_{tr}-\epsilon)\right]}{m(r_{tr})+4\pi r_{tr}^3 p},  
\end{equation}
with $r_{tr}$ and $\epsilon$ being the radial position where phase-transition occurs inside of the star and $\epsilon$ an infinitesimal parameter, respectively. Obviously, the region $r < r_{tr}$ represents the core of the star, and the region $r > r_{tr}$ depicts the envelope of the star.

\section{Equation of state and numerical method}\label{eos_employed}

\subsection{Equation of state}

\begin{figure}[ht]
\begin{center}
\includegraphics[width=0.98\linewidth]{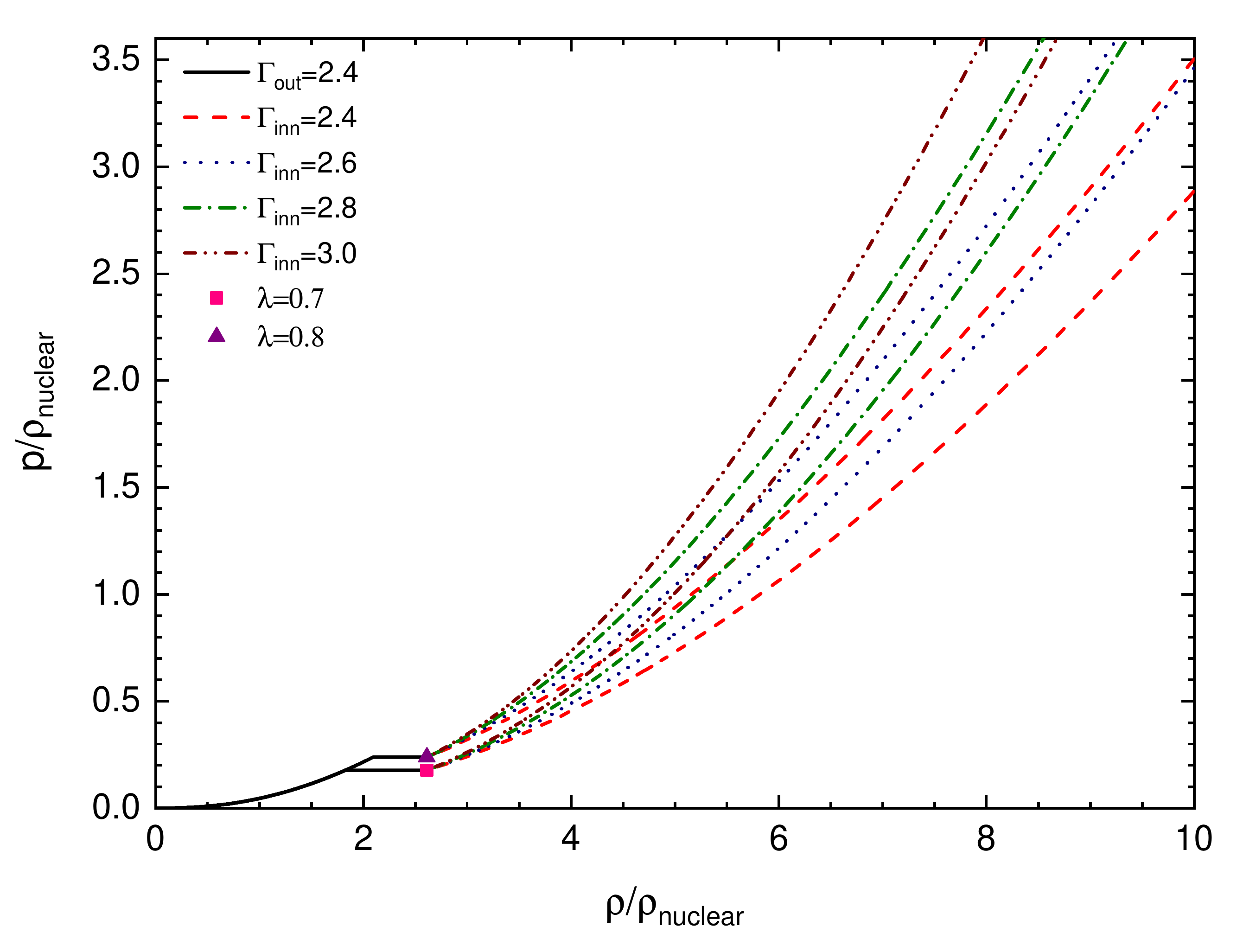}
\includegraphics[width=0.98\linewidth]{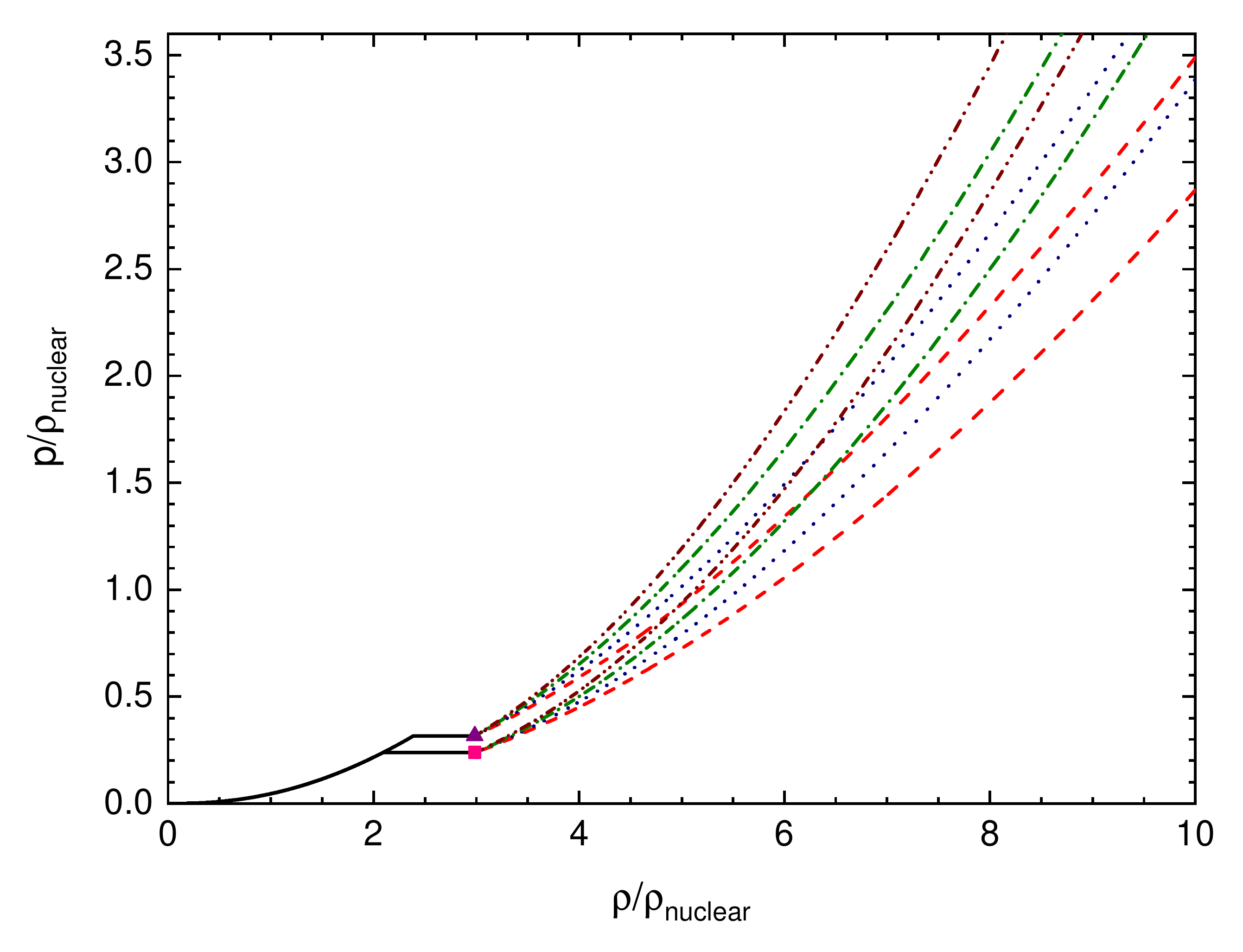}
\caption{Some equations of state for two density jump parameters $\lambda$. The fluid pressure $p$ and energy density $\rho$ are normalized by the nuclear density $\rho_{\rm nuclear}=2.68\times10^{17}\,[\rm kg/m^3]$. On the top and bottom panel are respectively adopted $\rho_{\rm inn}^{\rm dis}=7\times10^{17}\,[\rm kg/m^3]$ and $8\times10^{17}\,[\rm kg/m^3]$.}
\label{p_rhoc}
\end{center}
\end{figure}

To describe the matter that makes up the compact object, in the two-phase configurations, the relativistic polytropic equation of state \cite{tooper1965} is adopted. Then, the energy density and fluid pressure of each phase are respectively connected through the relations:
\begin{eqnarray}
\rho&=&\left(\frac{p}{K_{\rm inn}}\right)^{1/\Gamma_{\rm inn}}+\frac{p}{\Gamma_{\rm inn}-1},\quad p_{\rm inn}^{\rm dis}\leq p\leq p_c,\\
\rho&=&\left(\frac{p}{K_{\rm out}}\right)^{1/\Gamma_{\rm out}}+\frac{p}{\Gamma_{\rm out}-1}, \quad 0\leq p\leq p_{\rm out}^{\rm dis}.
\end{eqnarray}
These two relations bear parameters from the inner and outer regions denoted by the sub-indexes ``${\rm inn}$'' and ``${\rm out}$'', respectively; namely, $K_{\rm inn}$ and $K_{\rm out}$ are the polytropic constants, $\Gamma_{\rm inn}$ and $\Gamma_{\rm out}$ being the polytropic exponents, and $p^{\rm dis}_{\rm inn}$ and $p^{\rm dis}_{\rm out}$ represent the phase transition pressure. Following \cite{doneva2012}, we set the inner polytropic constant value: 
\begin{equation}
K_{\rm inn}=0.0195\times(1.67\times10^{17}\,{\rm kg/m^3})^{-1.34}.
\end{equation}
Since the fluid pressure should be continuous along the star, at the phase transition point, where the inner and outer phase transition pressures meet the condition $p_{\rm out}^{\rm dis}=p_{\rm inn}^{\rm dis}$, the outer polytropic constant takes the form:
\begin{equation}
K_{\rm out}=K_{\rm inn}\frac{\left(\rho_{\rm inn}^{\rm dis}-p_{\rm inn}^{\rm dis}/\left(\Gamma_{\rm inn}-1\right)\right)^{\Gamma_{\rm inn}}}{\left(\rho_{\rm out}^{\rm dis}-p_{\rm out}^{\rm dis}/\left(\Gamma_{\rm out}-1\right)\right)^{\Gamma_{\rm out}}}.
\end{equation}
At this point, the inner and outer phase transition energy densities are related by:
\begin{equation}
\rho_{\rm out}^{\rm dis}=\lambda\,\rho_{\rm inn}^{\rm dis},
\end{equation}
where $\lambda$ is known as the density jump parameter. Some examples of the EOS with a sharp density jump are presented in Fig. \ref{p_rhoc}.

In the next section, we consider the value of $\Gamma_{\rm inn}=2.4$, $\Gamma_{\rm inn}=2.6$ and $\Gamma_{\rm out}=2.4$ and the phase transition parameter $\lambda$ between the values $0.5$ and $1.0$, in order to analyze both the role of a stiffer core fluid and the effects of a more abrupt phase transition in the star equilibrium configuration, respectively. The values of $\Gamma_{\rm inn}, \Gamma_{\rm out}$ and $\lambda$ chosen allow us to obtain comparable results with the data found through observation, which are reported by the LIGO-Virgo network in \cite {abbott_2018a_tidal} and by NICER in \cite{riley2019,miller2019,riley2021,miller2021,cromartie2020,antoniadis2013,demorest2010}.


\subsection{Numerical method}

The effects of the phase transition on the equilibrium configurations and tidal deformations are investigated through the numerical solution of the system of equations, boundary conditions, and junction conditions established in the section \ref{sec-basicequations}, for each $\rho_{\rm inn}^{\rm dis}$, $\lambda$, $\Gamma_{\rm inn}$, and $\Gamma_{\rm out}$. This system of equations is integrated from the center toward the star's surface.

The analysis of the radial stability starts by solving the stellar structure equations by using the Runge-Kutta fourth-order method in order to determine the radial pulsation coefficients. After, we begin at the star's core with the solution of Eqs. \eqref{ro1}-\eqref{ro3} for a test value of $\omega^2$. These equations are numerically integrated outwards until the interface is found, where the junction conditions are employed (for the slow case Eqs. \eqref{slow_condition_1} and \eqref{slow_condition_2} and for the rapid case Eqs. \eqref{rapid_condition_1} and \eqref{rapid_condition_2}) to find the values of $\xi$ and $\Delta p$ at the other side of the interface. Then, the numerical integration continues towards the surface of the stars attempting to reach the conditions \eqref{null_pressure} and \eqref{delta_p}. Whether, after each integration, the equality \eqref{delta_p} is not fulfilled, $\omega^2$ is corrected until satisfying this equality in the next integration. The parameters $\omega^2$ which satisfy the oscillation are called eigenvalues of the radial pulsation equation and $\omega$ of eigenfrequencies (review \cite{pereira_flores2018}).

We are interested in analyzing the radial stability of stars, we only analyze the lowest eigenvalue, i.e., $\omega_{0}^{2}$. When $\omega_{0}^{2}>0$, the star is stable against small radial perturbations. $\omega_0$ is known as the eigenfrequency of the fundamental mode.

\section{Results}\label{results}

\subsection{Equilibrium configuration of neutron stars}\label{results1}

\begin{figure}[ht]
\begin{center}
\includegraphics[width=0.98\linewidth]{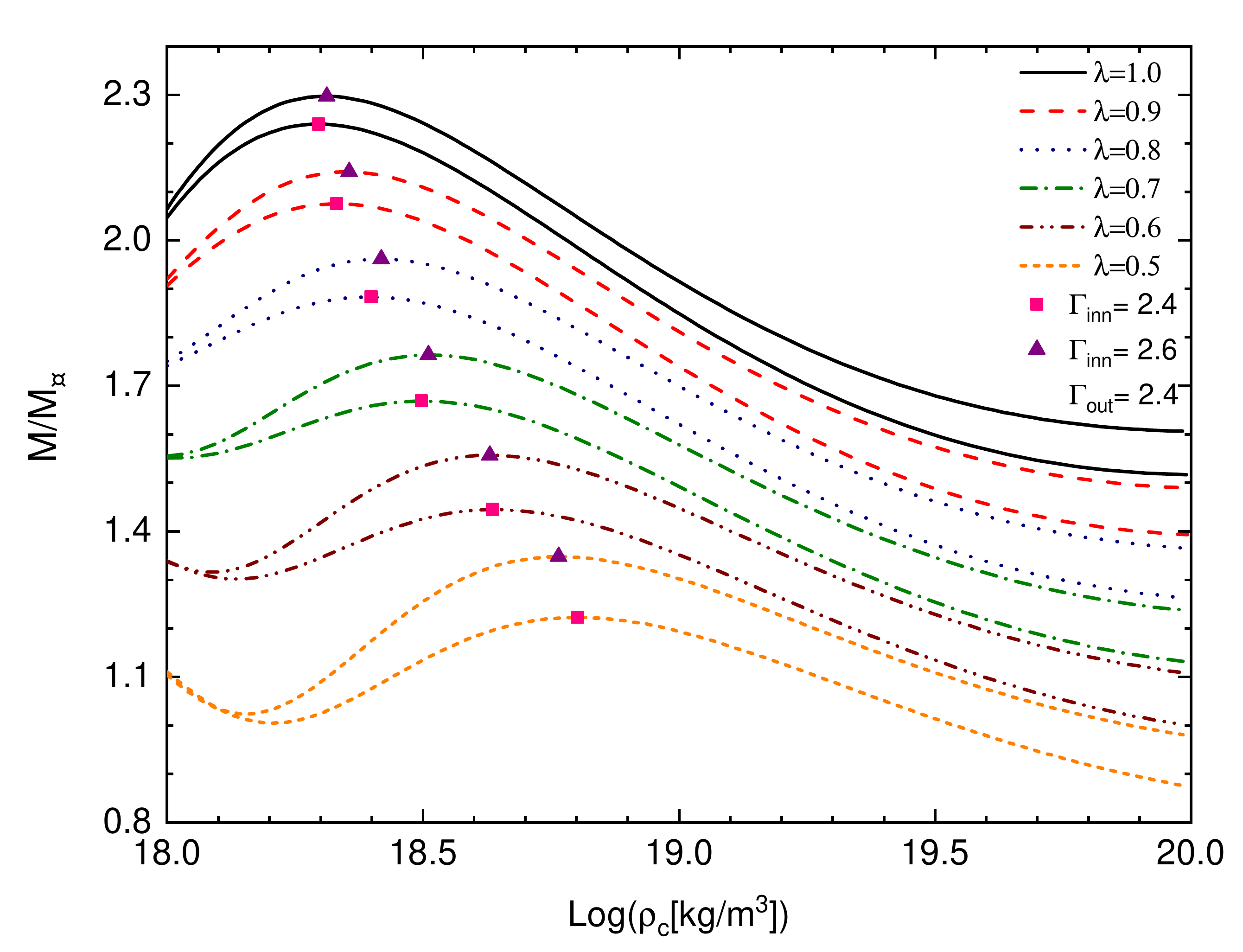}
\caption{Mass against the central energy density for different values of $\lambda$, $\rho_{\rm inn}^{\rm dis}=8\times10^{17}\,[\rm kg/m^3]$, $\Gamma_{\rm out}=2.4$, and two different values of $\Gamma_{\rm inn}$. }
\label{M_rhoc}
\end{center}
\end{figure}

The compact star total mass sequence, normalized to the Sun’s mass $M_{\odot}$, as a function of the central energy density is plotted in Fig. \ref{M_rhoc} for two values of $\Gamma_{\rm inn}$, four different values of $\lambda$, $\rho_{\rm inn}^{\rm dis}=8\times10^{17}\,[\rm kg/m^3]$, and $\Gamma_{\rm out}=2.4$. The central energy density goes from $10^{18}$ to $10^{20}\,[\rm kg/m^3]$. On the panel, the total mass increases with the central energy density until reaches the maximum mass of the sequence, after this point, $M/M_{\odot}$ decreases monotonically with the increment of $\rho_c$.

In the panel, it is noted the diminution of the total mass with the density jump parameter. This is associated with the fact that the fluid pressure decays abruptly due to the presence of a phase transition, being this declines greater for lower $\lambda$ (see Fig. \ref{p_rhoc}). The change of the mass with $\Gamma_{\rm inn}$ is also observed in Fig. \ref{M_rhoc}. For a fixed $\rho_c$ and $\lambda$, for a greater $\Gamma_{\rm inn}$, a larger total mass is derived. This point can be understood since for larger interior polytropic exponents larger central pressures are obtained. Thus, larger central pressure supports more mass against gravitational collapse. In addition, it is important to say that, for larger $\Gamma_{\rm inn}$, neutron stars with more compact cores are obtained (review, e.g., \cite{alz-poli-qbh,alz-2eos-qbh}).

\begin{figure}[ht]
\begin{center}
\includegraphics[width=0.98\linewidth]{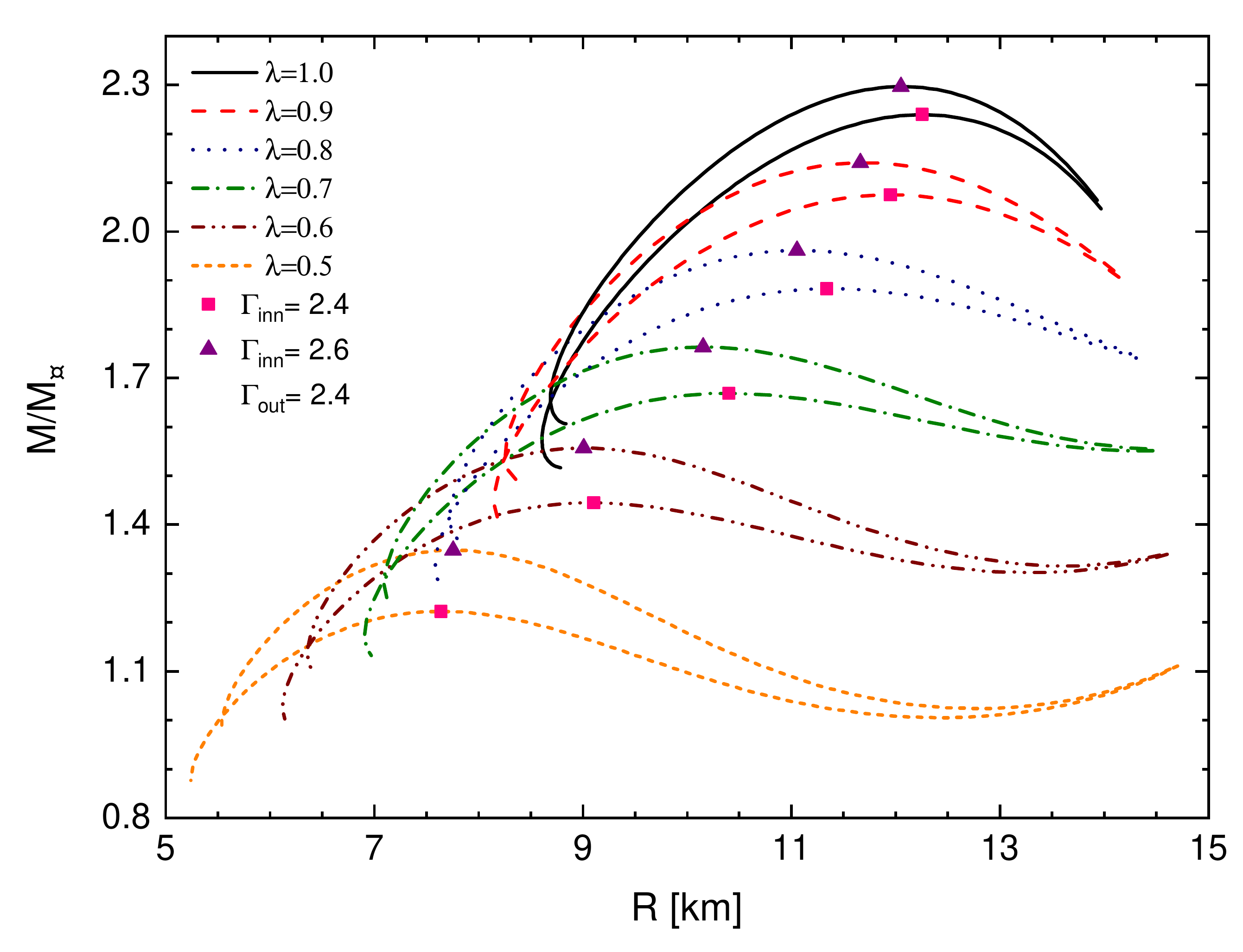}
\caption{Total mass as a function of the radius for four values of $\lambda$, $\rho_{\rm inn}^{\rm dis}=8\times10^{17}\,[\rm kg/m^3]$, $\Gamma_{\rm out}=2.4$, and two different values of $\Gamma_{\rm inn}$.} 
\label{r_m}
\end{center}
\end{figure}

Fig. \ref{r_m} shows the mass as a function of the total radius for two values of $\Gamma_{\rm inn}$, for $\rho_{\rm inn}^{\rm dis}=8\times10^{17}[\rm kg/m^3]$, few values of $\lambda$, and $\Gamma_{\rm out}=2.4$. As in the case of Fig. \ref{M_rhoc}, in this figure is considered $\rho_c$ between $10^{18}$ and  $10^{20}\,[\rm kg/m^3]$. The panel shows an increment of $M/M_{\odot}$ with the diminution of $R$ until to find a $M_{\rm max}/M_{\odot}$. After this point, the curves turn anti-clockwise, to $M(R)$ starts to decrease with $R$ until to reach $R_{\rm min}$. From here on, the mass decays with the increment of the total radius.

In Fig. \ref{r_m}, for some range of central energy density, we also find an increment of the total radius with the diminution of the density jump parameter. This is due to the fact that in these compact stars, the pressure of the fluid decays slower with the increase of the radial coordinate, thus obtaining larger radii. In addition, for some range of $\rho_c$ and $\lambda$, we can also observe decrements of the total radius with the increment of $\Gamma_{\rm inn}$. Despite having an increase of the central pressure with $\Gamma_{\rm inn}$, it decays faster with the growth of the radial coordinate. In this way, compact objects have a smaller total radius. Finally, in Fig. \ref{r_m}, we also see that the radius of the stars with maximum mass decreases with the jump in phase transition density. This is due to the fact that the central pressure of the star decreases with $\lambda$, thus, the pressure decays faster with the radial coordinate.

\begin{figure}[ht]
\begin{center}
\includegraphics[width=0.98\linewidth]{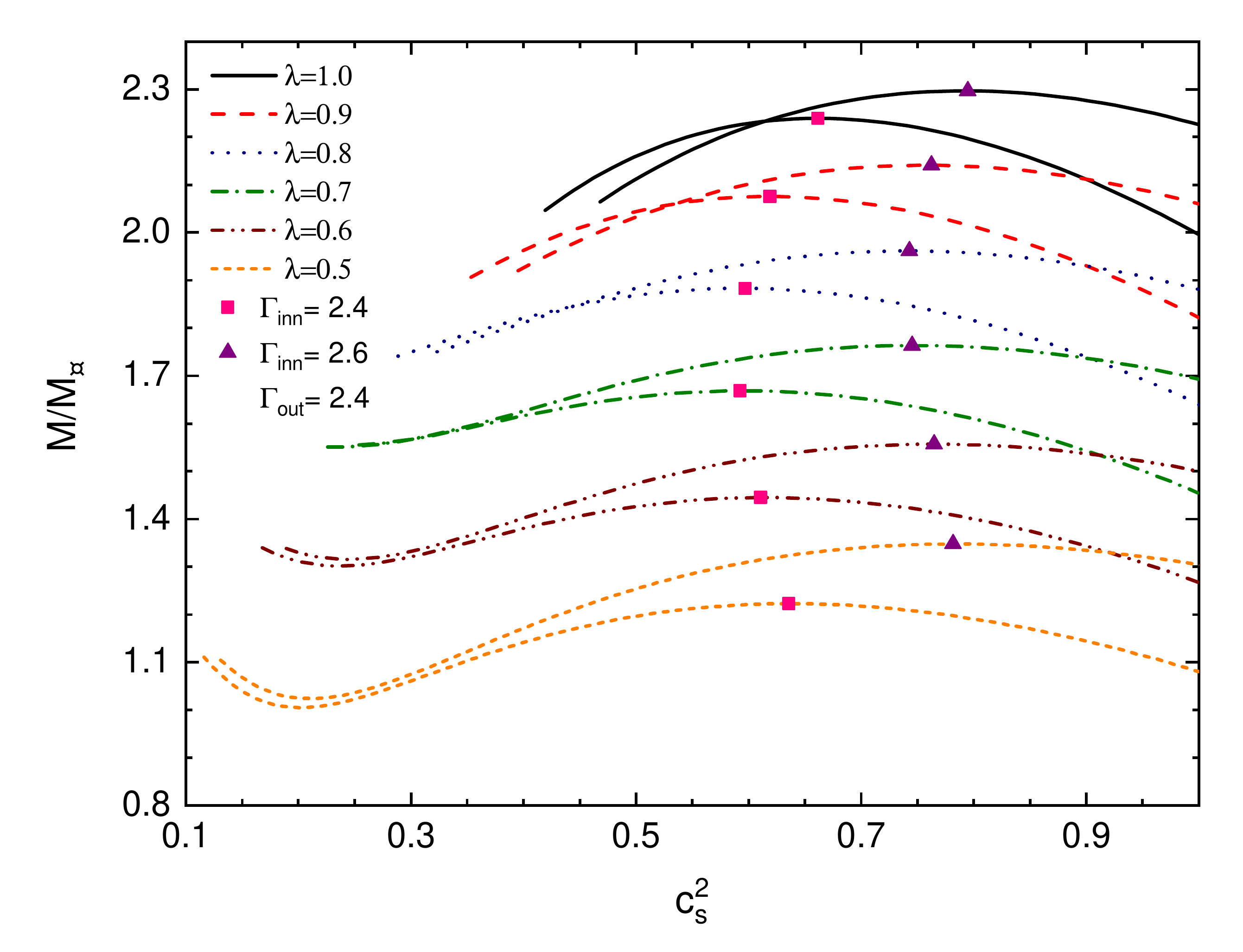}
\caption{Total mass against the speed of sound for four values of $\lambda$, $\rho_{\rm inn}^{\rm dis}=8\times10^{17}\,[\rm kg/m^3]$, $\Gamma_{\rm out}=2.4$, and two different values of $\Gamma_{\rm inn}$.} 
\label{m_cs2}
\end{center}
\end{figure}

The total mass versus the speed of sound is plotted in Fig. \ref{m_cs2}, for $\rho_{\rm inn}^{\rm dis}=8\times10^{17}[\rm kg/m^3]$, some values of $\lambda$, and $\Gamma_{\rm out}=2.4$. In this figure, we only present equilibrium configurations with the speed of sound lower than the speed of light  $c_s^2=1.0$. As can be seen, in particular, at stars that present first-order phase transition with low central energy densities (stars with low total masses), the speed of sound never exceeds the conformal limit $c_s^2=1/3$. However, in stars with larger total masses, the speed of sound exceeds the conformal limit value but is far to attain the speed of light. These results are in concordance with those reported in the article \cite{annala_2022}.

\subsection{Radial stability of neutron stars}

\begin{figure*}[ht]
\begin{center}
\includegraphics[width=0.45\linewidth]{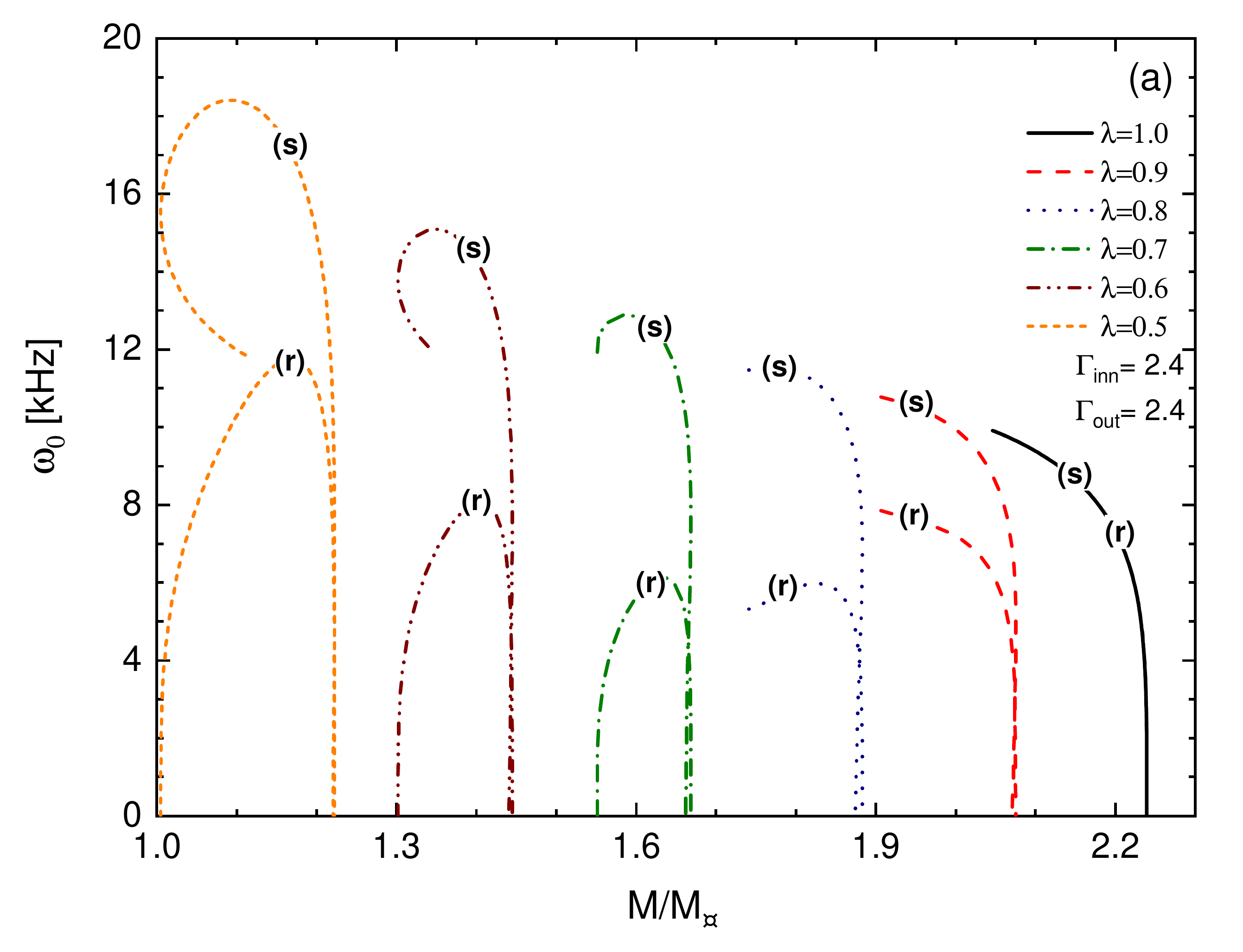}
\includegraphics[width=0.45\linewidth]{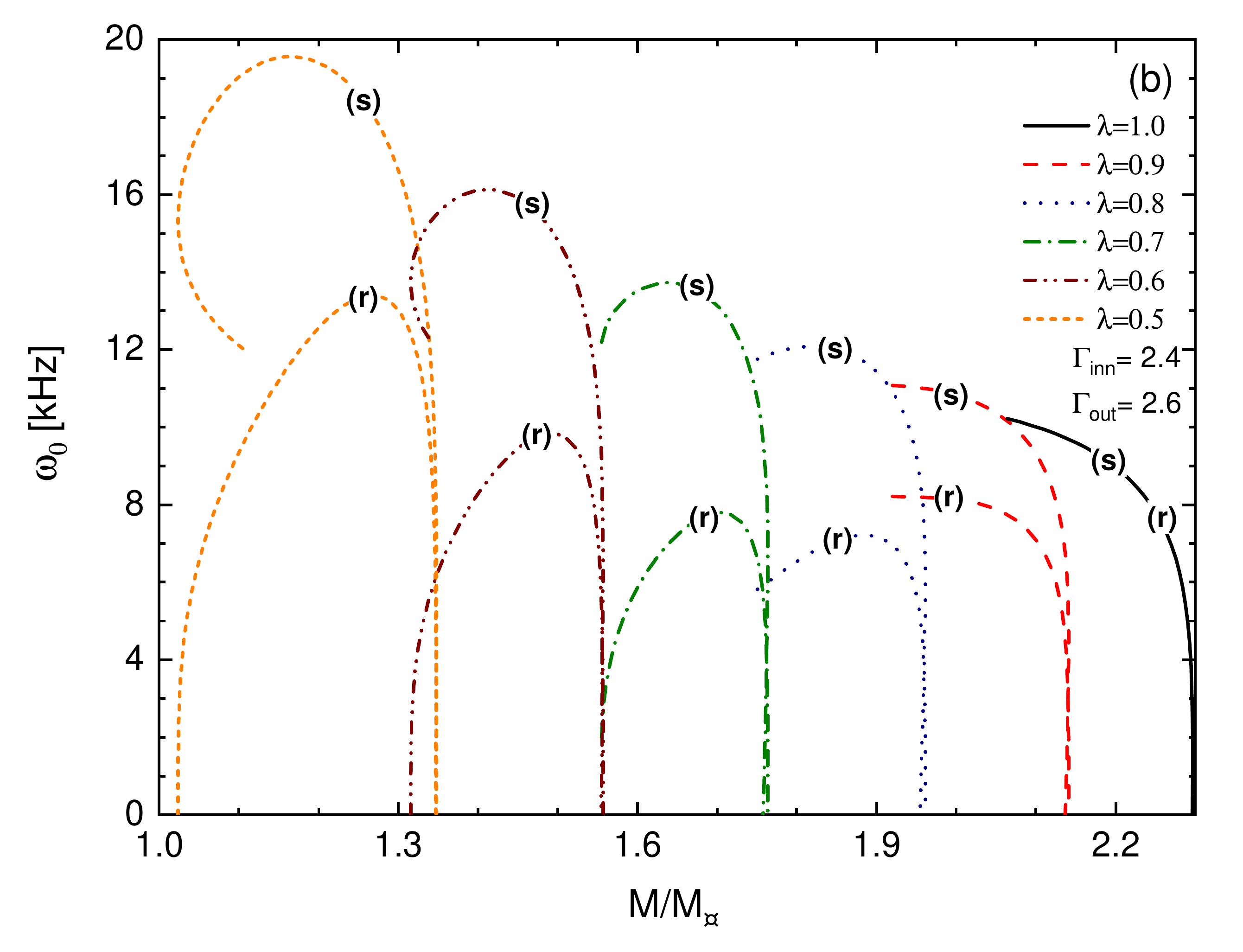}
\includegraphics[width=0.45\linewidth]{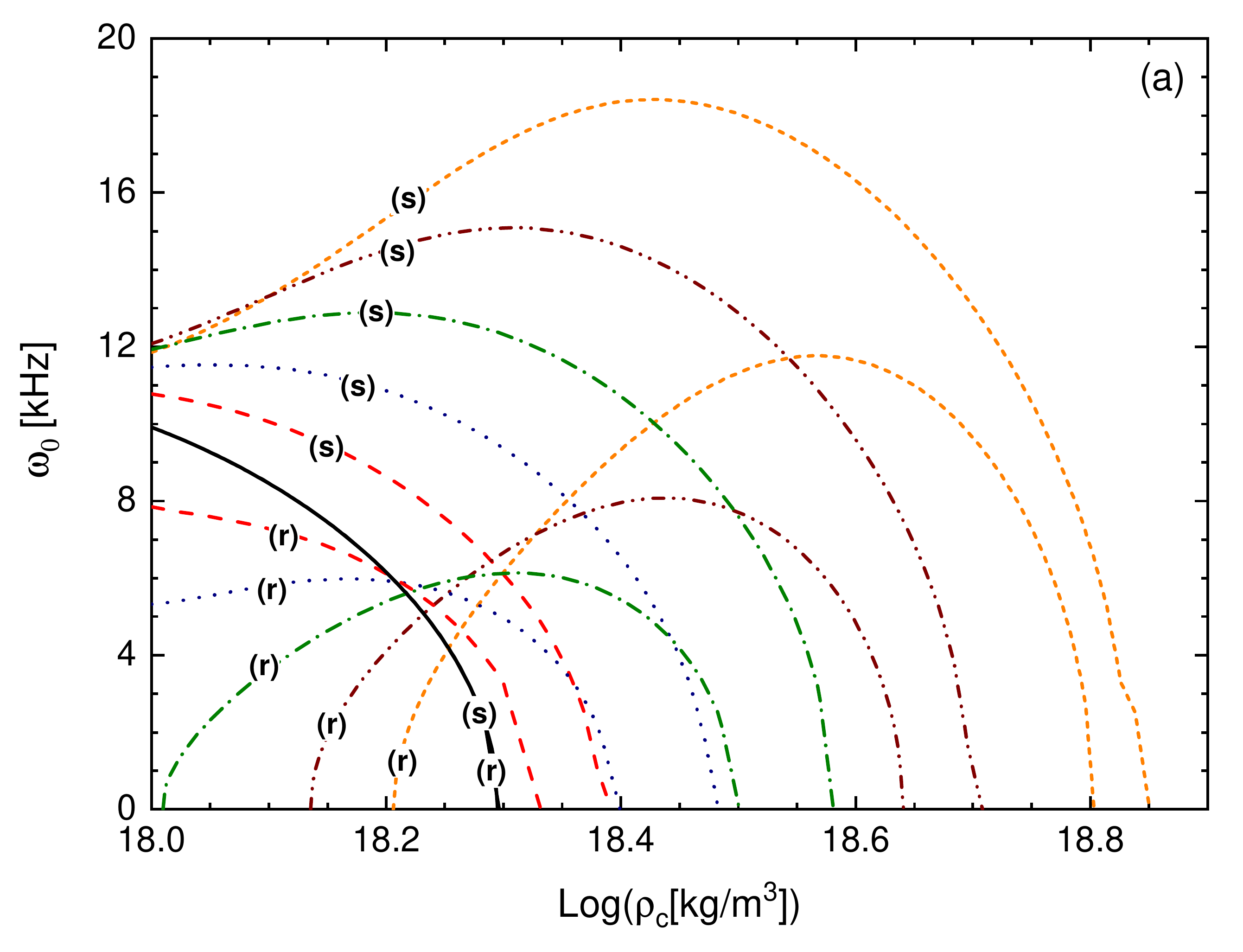}
\includegraphics[width=0.45\linewidth]{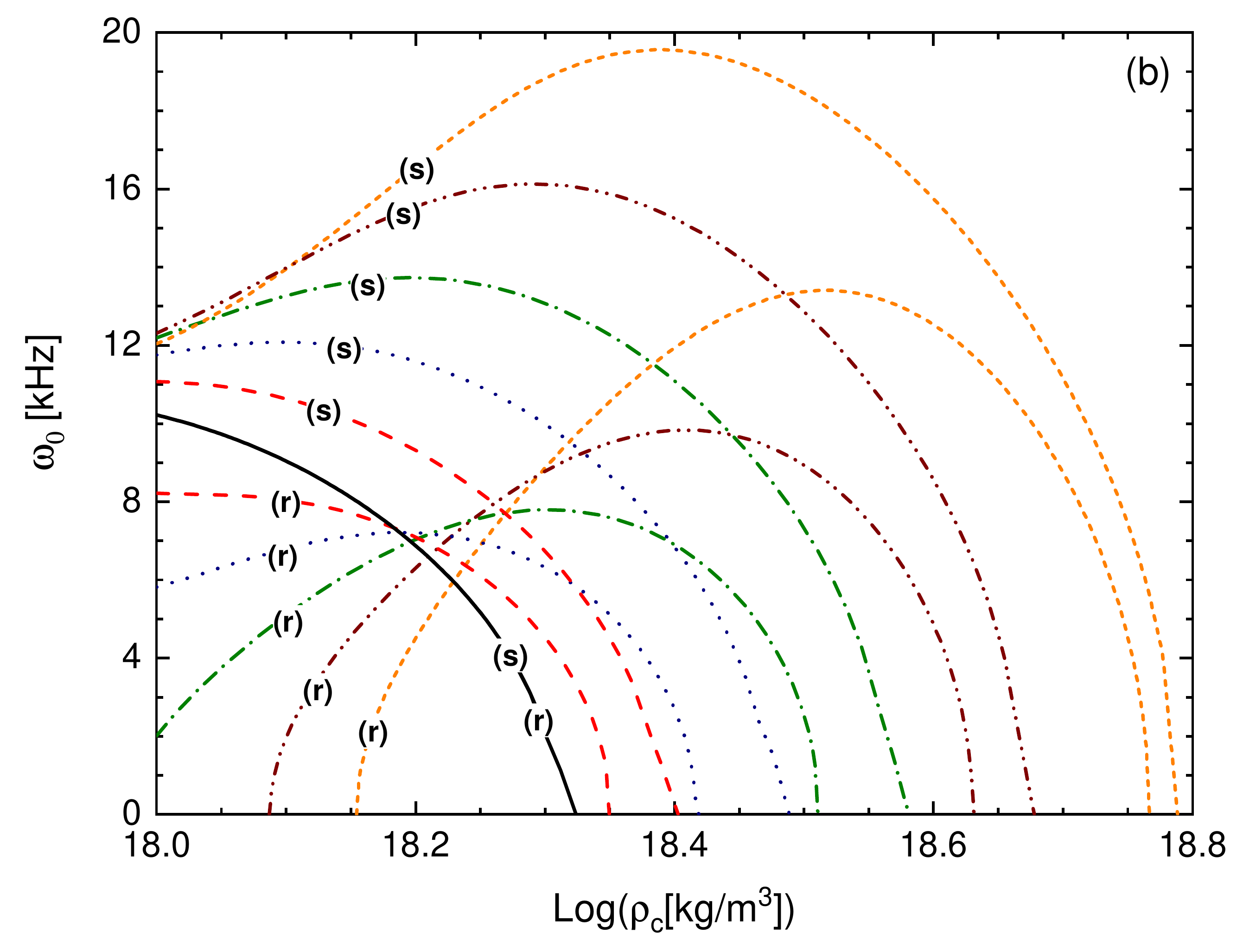}
\caption{The slow $(s)$ and rapid $(r)$ eigenfrequency of the fundamental mode $\omega_0$ as a function of the total mass $M/M_{\odot}$ and against the central energy density for $\rho_{\rm inn}^{\rm dis}=8\times10^{17}\,[\rm kg/m^3]$, $\Gamma_{\rm out}=2.4$, four values of $\lambda$, and two values of $\Gamma_{\rm inn}$. On the left panels, it is used $\Gamma_{\rm inn}=2.4$ and on the right panels, it is employed $\Gamma_{\rm inn}=2.6$.} 
\label{RS_omega_masa}
\end{center}
\end{figure*}

Fig. \ref{RS_omega_masa} shows the behavior of the slow and rapid eigenfrequency of the fundamental oscillations as a function of the total mass and against the central energy density for some different values of $\lambda$, $\rho_{\rm inn}^{\rm dis}=8\times10^{17}\,[\rm kg/m^3]$, $\Gamma_{\rm out}=2.4$, and $\Gamma_{\rm inn}=2.4$ on the left panels, and $\Gamma_{\rm inn}=2.6$ on the right panels. From the figure, for $\lambda=1$, it can be noted that the maximum total mass is found at the zero eigenfrequencies of oscillation. This case represents the usual study of radial oscillation of neutron stars in the absence of phase transition. In turn, for $\lambda<1$, as well as in \cite{pereira_flores2018}, we note that the total mass at the null eigenfrequency of oscillation depends on the type of the phase transition. At this point, when $\Gamma_{\rm inn}=\Gamma_{\rm out}=2.4$, the difference of the mass attained in the rapid and slow case is almost $0.246\%$ for $\lambda=0.9$, $0.470\%$ for $\lambda=0.8$, $0.397\%$ for $\lambda=0.7$, $0.299\%$ for $\lambda=0.6$, and $0.164\%$ for $\lambda=0.5$. In the case $\Gamma_{\rm inn}=2.6$ and $\Gamma_{\rm out}=2.4$, the difference of the mass is around $0.197\%$ for $\lambda=0.9$, $0.358\%$ for $\lambda=0.8$, $0.330\%$ for $\lambda=0.7$, $0.161\%$ for $\lambda=0.6$, and $0.0371\%$ for $\lambda=0.5$ (review Table \ref{tabla:stability}). 

In Fig. \ref{RS_omega_masa}, in the slow case, the total mass at the zero eigenfrequencies of oscillation is derived at $\rho_c$ larger than the one employed to obtain the maximum mass value; i.e., twins stars are derived, stable stars with the same total mass but with both different central energy densities and total radii. However, in the rapid case, the maximum mass and the null eigenfrequency of oscillation are obtained by using the same value of $\rho_c$. It indicates that in a sequence of static equilibrium configurations, the maximum mass point marks the beginning of the instability against small radial perturbations. This characteristic in each phase transition could be useful to differentiate them.

\begin{table*}[ht!]
\centering
\begin{tabular}{|c|c|c|c|c|c||c|c|c|c|}
\cline{3-10}
\multicolumn{2}{c}{}&\multicolumn{4}{|c||}{$\omega_{0,{\rm s}}=0$} & \multicolumn{4}{c|}{$\omega_{0,{\rm r}}=0$}\\\cline{1-10} $\Gamma_{\rm inn}$ &  $\lambda$ & $\rho_c\,[\rm kg/m^3]$          & $M/M_{\odot}$ & $R\,[\rm km]$& $R_{\rm core}\,[\rm km]$ & $\rho_c\,[\rm kg/m^3]$             & $M/M_{\odot}$ & $R\,[\rm km]$& $R_{\rm core}\,[\rm km]$
\\\cline{1-10}
 & $1.0$ & $1.9776\times10^{18}$ & $2.2391$ & $12.254$& $12.254$ & $1.9776\times10^{18}$ & $2.2391$ & $12.254$& $12.254$\\\cline{2-10}
 & $0.9$ & $2.5095\times10^{18}$ & $2.0689$& $11.525$& $7.8405$& $2.1462\times10^{18}$& $2.0753$& $11.949$& $7.8218$\\\cline{2-10}
\multirow{3}{0.8cm}{$2.4$} & $0.8$ & $3.0423\times10^{18}$ & $1.8742$& $10.800$& $7.5842$& $2.5079\times10^{18}$& $1.8830$& $11.339$& $7.6380$\\\cline{2-10}
 & $0.7$ & $3.8103\times10^{18}$ & $1.6619$& $9.8790$& $7.1888$& $3.1426\times10^{18}$&  $1.6685$& $10.403$& $7.3066$\\\cline{2-10}
 & $0.6$ & $5.1074\times10^{18}$ & $1.4405$& $8.7010$& $6.6230$& $4.3246\times10^{18}$&  $1.4448$& $9.1068$& $6.7657$\\\cline{2-10}
 & $0.5$ & $7.0873\times10^{18}$ & $1.2207$& $7.4132$& $5.9219$& $6.3371\times10^{18}$&  $1.2227$& $7.6410$& $6.0283$\\
 \hline \hline
 & $1.0$ & $2.0514\times10^{18}$ & $2.2966$& $12.050$& $12.050$& $2.0514\times10^{18}$ & $2.2966$& $12.050$& $12.050$\\\cline{2-10}
 & $0.9$ & $2.5269\times10^{18}$ & $2.1368$& $11.375$& $8.0505$& $2.2691\times10^{18}$& $2.1410$& $11.661$& $8.0615$\\\cline{2-10}
\multirow{3}{0.8cm}{$2.6$} & $0.8$ & $3.0803\times10^{18}$ & $1.9542$& $10.623$& $7.7841$& $2.6228\times10^{18}$& $1.9612$& $11.054$& $7.8585$\\\cline{2-10}
 & $0.7$ & $3.8087\times10^{18}$ & $1.7576$& $9.7431$& $7.3940$& $3.2424\times10^{18}$&  $1.7634$& $10.155$& $7.5138$\\\cline{2-10}
 & $0.6$ & $4.7597\times10^{18}$ & $1.5543$& $8.7586$& $6.8949$& $4.2734\times10^{18}$&  $1.5568$& $9.0081$& $6.9936$\\\cline{2-10}
 & $0.5$ & $6.1448\times10^{18}$ & $1.3472$& $7.6503$& $6.2731$& $5.8217\times10^{18}$&  $1.3477$& $7.7576$& $6.3259$\\
 \hline 
\end{tabular}
\caption{\label{tabla:stability}The central energy density $\rho_c$, the total mass $M/M_{\odot}$, the total radius $R$, and the core radius $R_{\rm core}$ where the null eigenfrequency of oscillation of the slow case $\omega_{0,s}$ and rapid case $\omega_{0,r}$ are derived. These parameter are found for $\rho_{\rm inn}^{\rm dis}=8\times10^{17}\,[\rm kg/m^3]$, $\Gamma_{\rm out}=2.4$ and different values of $\Gamma_{\rm inn}$ and $\lambda$.}
\end{table*}

Table \ref{tabla:stability} presents the central energy densities and total masses where the zero eigenfrequencies of oscillations for the slow and the rapid phase transition. These parameters are derived for $\rho_{\rm inn}^{\rm dis}=8\times10^{17}\,[\rm kg/m^3]$, $\Gamma_{\rm out}=2.4$ and some values of $\Gamma_{\rm inn}$ and $\lambda$. In the table, for a fixed $\Gamma_{\rm inn}$, at the null eigenfrequency of oscillation for the slow and rapid case, we note that the total mass, the total radius, and the core radius decrease with the density jump parameter. This could be understood since $p_c$ decays with $\lambda$, in this way, the total pressure diminishes faster with the growth of the radial coordinate. On the other hand, when $\Gamma_{\rm inn}$ is increased, i.e., when a stiffer core fluid is considered, stars with a core radius $R_{\rm core}$ closer to the total radius $R$ are found.

\subsection{Tidal deformability in the light of GW$170817$}

\begin{figure}[ht]
\begin{center}
\includegraphics[width=1.0\linewidth]{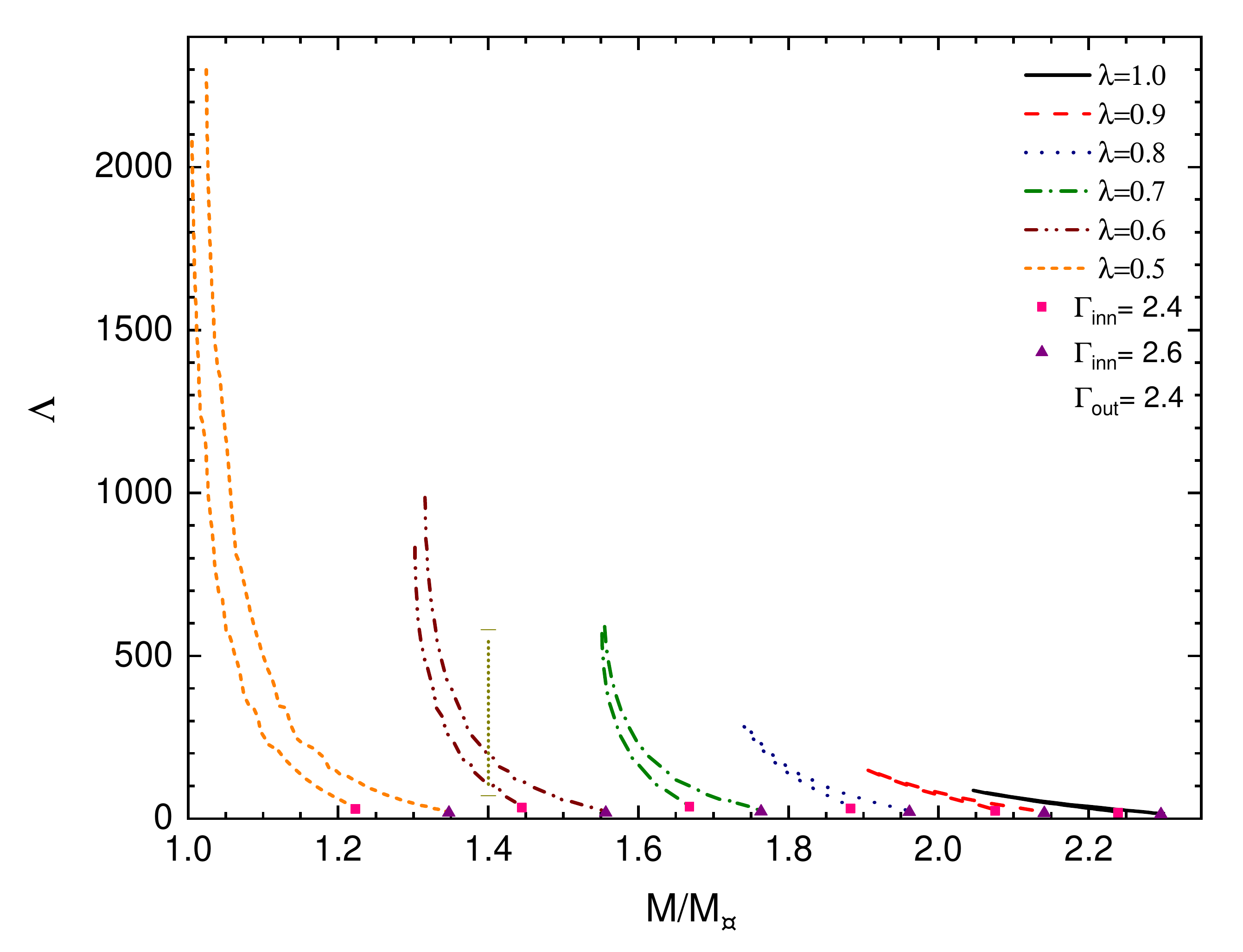}
\includegraphics[width=1.0\linewidth]{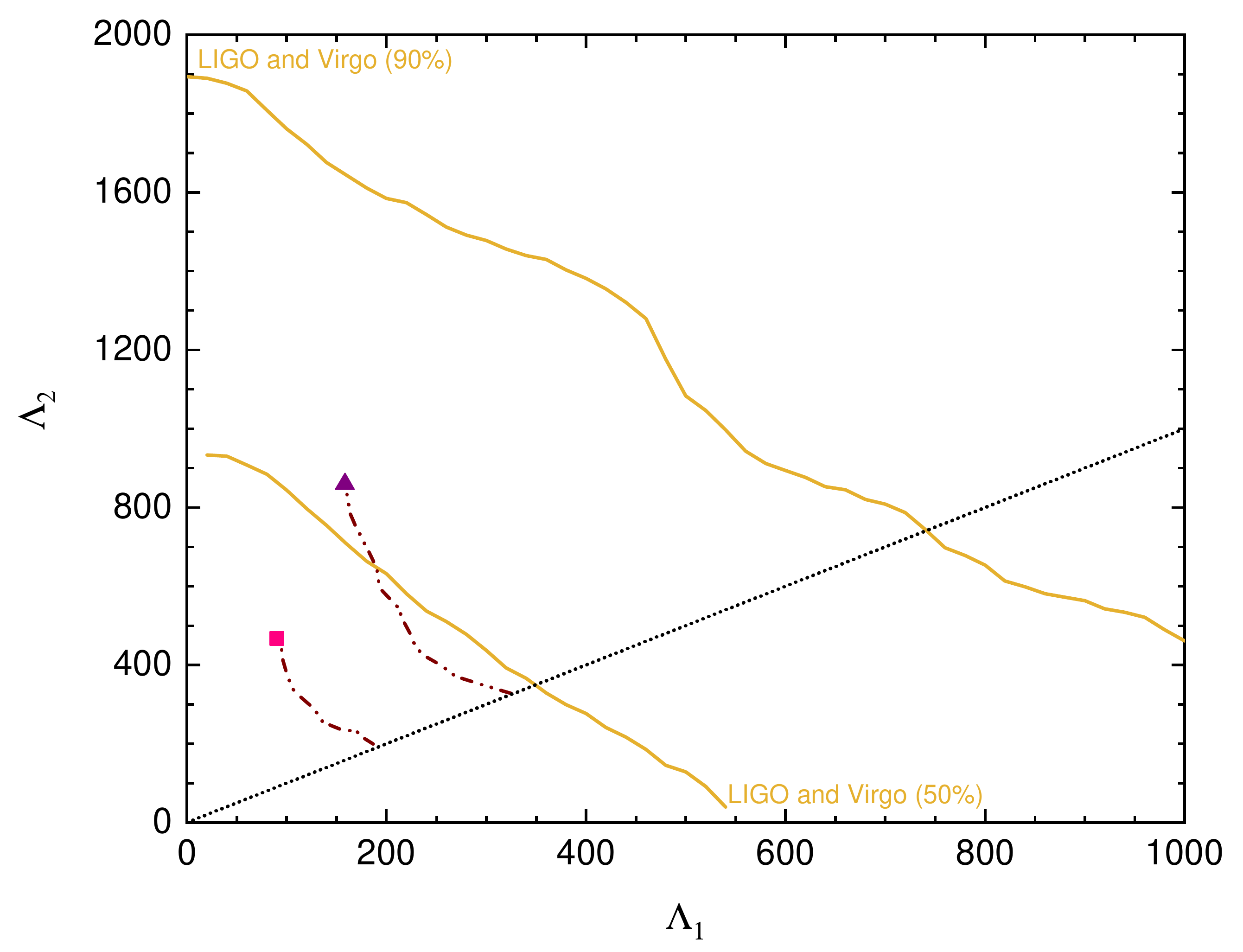}
\caption{Top: Dimensionless tidal deformability $\Lambda$ as a function of the total mass in Solar masses. The vertical dash-dot-dot line marks the tidal deformability of event GW$170817$ estimated in \cite{abbott_2018a_tidal}. Bottom: The dimensionless tidal deformability $\Lambda_1$ and $\Lambda_2$ for a binary NS system with masses $m_1$ and $m_2$ and the same chirp mass as the event GW$170817$ \cite{abbott2017_NS}. We only take into account the combination with $m_1>m_2$. The diagonal short dot line denotes the $\Lambda_1=\Lambda_2$ limit. The solid top yellow line indicates the $90\%$ credibility level and the bottom yellow solid line is the $50\%$ level established by LIGO-Virgo scientific network in the low-spin prior scenario. Only stable equilibrium configurations with slow conversions at the interface are shown.} 
\label{TD_masa}
\end{center}
\end{figure}

Tidal deformability against the total mass of stable NS is plotted at the top panel of Fig. \ref{TD_masa} for $\rho_{\rm inn}^{\rm dis}=8\times10^{17}\,[\rm kg/m^3]$, $\Gamma_{\rm out}=2.4$, different values of $\lambda$ and $\Gamma_{\rm inn}$. On the panel, it is also presented the tidal deformability constrained by the event GW$170817$ for a star of $1.4M_{\odot}$ to be $70\leq\Lambda_{1.4M_{\odot}}\leq580$ \cite{abbott_2018a_tidal}, at 90\% confidence level, for low-spin priors. 

For a fixed $\lambda$ and masses range, at the top of Fig. \ref{TD_masa}, we note an increment of the tidal deformability when a stiffer fluid is considered in the core. On the other hand, by setting the parameter $\Gamma_{\rm inn}$ for different values of the phase transition parameter $\lambda$, we can notice that the importance of the parameter $\Lambda$ changes in a more relevant way. The lower the value of $\lambda$, the less the pressure of transition (see Fig. \ref{p_rhoc}) and the smaller the value of the deformability for the same mass. From these results, the effect of phase transition and stiffer fluid in the core are noticeable in the tidal deformability. Nonetheless, between these two factors, the phase transition parameter affects the NS properties more.

At the bottom panel of Fig. \ref{TD_masa}, the curves $\Lambda_1-\Lambda_2$ are presented for a binary NS system with chirp mass equal to GW$170817$. Since the total mass is associated with the dimensionless tidal deformability (top panel of Fig. \ref{TD_masa}), the curves $\Lambda_1-\Lambda_2$ are obtained once chosen a value of $m_1$ and calculating $m_2$ for the fixed value of the chirp mass ${\cal M}=1.188\,M_{\odot}$ \cite{abbott2017_NS}, which is theoretically calculated by the relation:
\begin{equation}\label{m1m2}
{\cal M}=\frac{(m_1\,m_2)^{3/5}}{(m_1+m_2)^{1/5}}.
\end{equation}
The values considered for $m_1$ and $m_2$ run from $1.36M_{\odot}\leq m_1\leq 1.60M_{\odot}$ and $1.17M_{\odot}\leq m_2\leq 1.36M_{\odot}$, respectively. 

At the bottom panel of Fig. \ref{TD_masa}, we investigate the effects of the phase transition and a stiffer fluid in the core (for $\rho_{\rm inn}^{\rm dis}=8\times10^{17}\,[\rm kg/m^3]$) of two neutron stars in a binary system. In this case, we note that both the phase transition and a stiffer equation of state could play an important role in the detection of these compact objects. From the results, we note that there is an interval for the density jump parameter $\lambda$, $0.7\leq\lambda \leq 0.8$, which plays the compact stars inside of $50\%$ and $90\%$ regions. The compact stars with $\lambda \geq 0.9$ are outside the $90\%$ region and do not appear in the panel. This is realized with the aim that the curves shown inside $50\%$ and $90\%$ can be seen clearly. Furthermore, we observe that for smaller values of $\lambda$ the curve change to smaller values of dimensionless deformability. On the other hand, it can be noted that a stiffer fluid in the core produces larger values of deformability.

\subsection{Change of stellar physical parameters with phase transition energy density and its comparison with observational data}

\begin{figure}[ht]
\begin{center}
\includegraphics[width=0.98\linewidth]{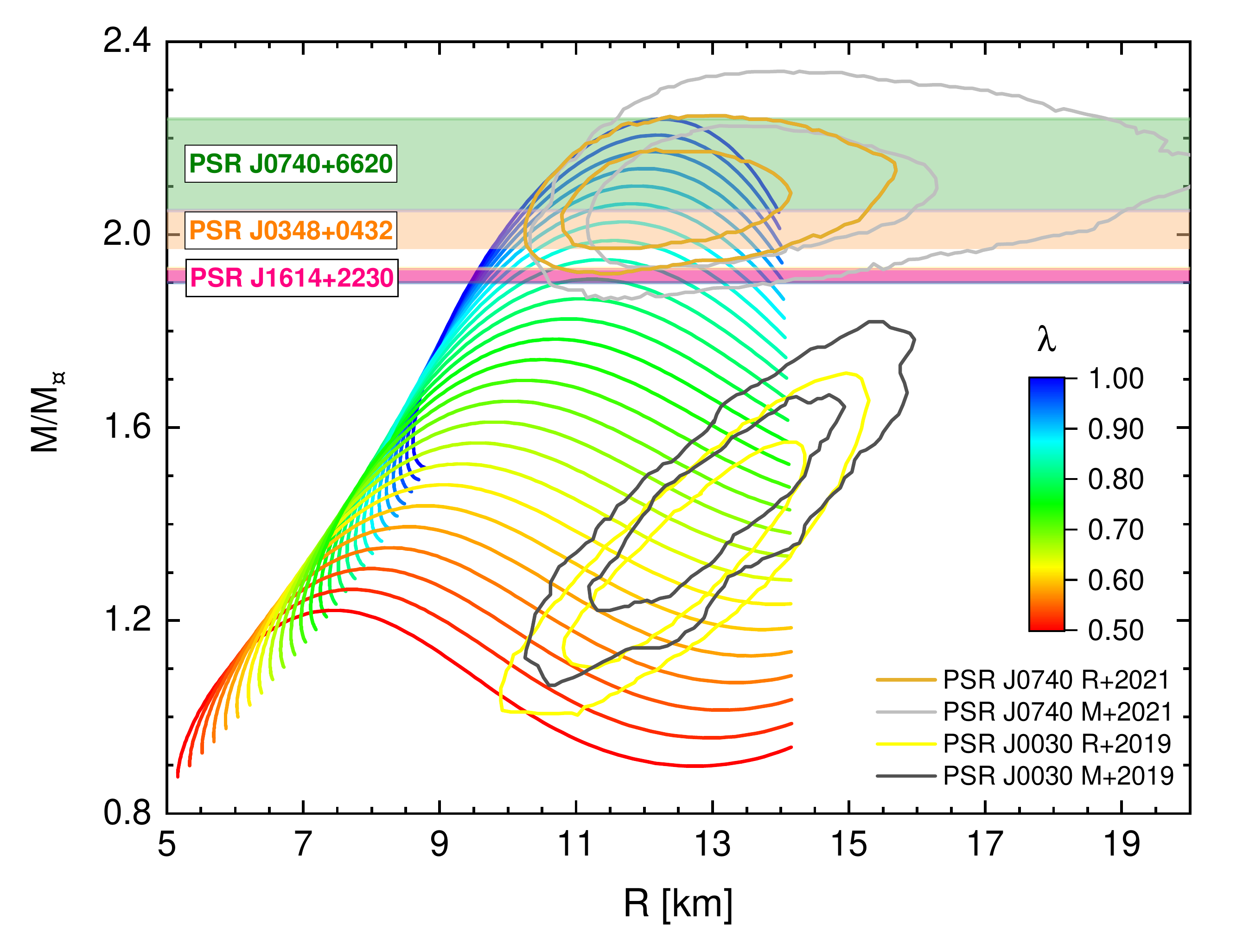}
\includegraphics[width=0.98\linewidth]{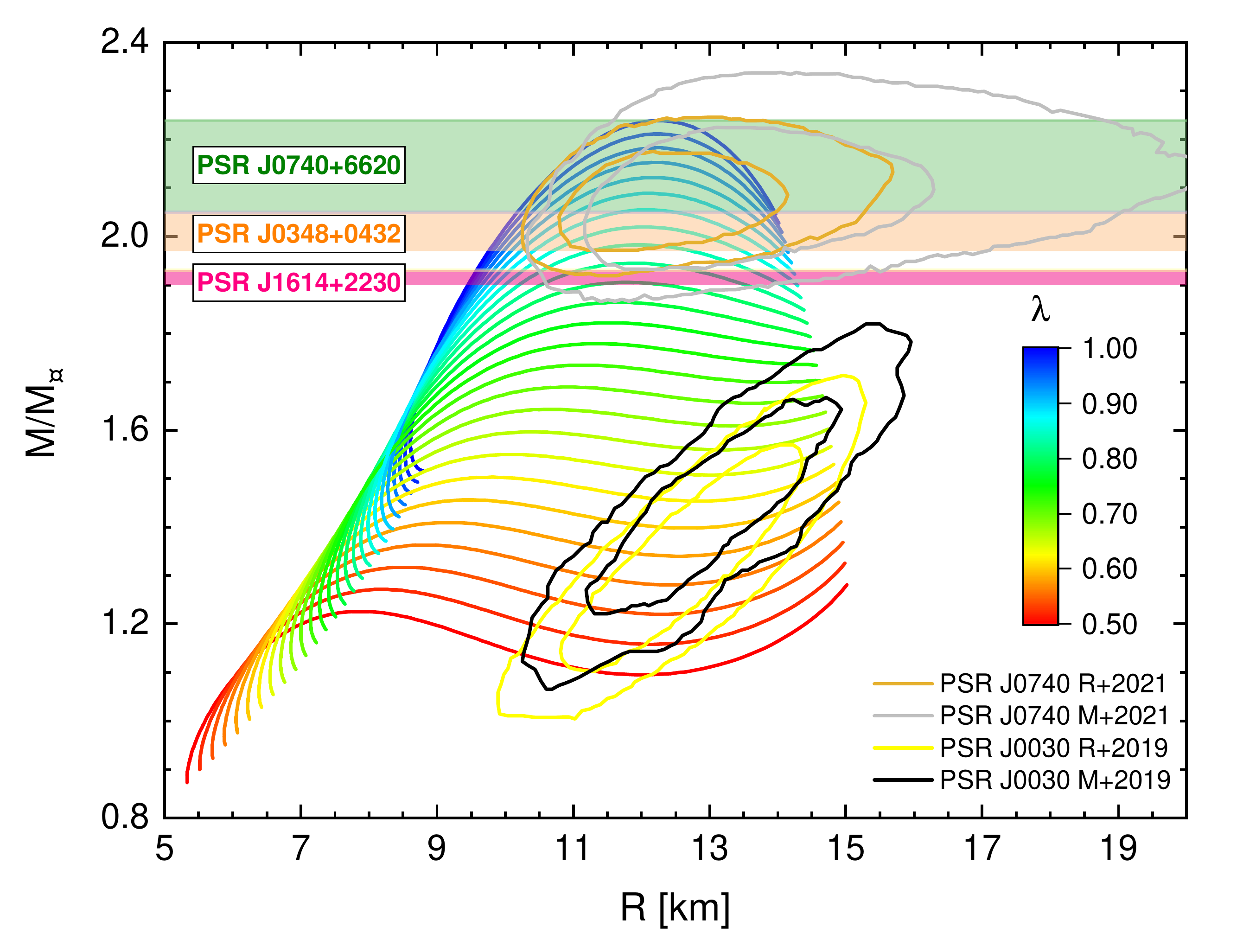}
\caption{Comparison of mass-radius curves with observational data using different values of $\lambda$, $\Gamma_{\rm inn}=\Gamma_{\rm out}=2.4$ and two different values of $\rho_{\rm inn}^{\rm dis}$. On the top and bottom panel are respectively used $\rho_{\rm inn}^{\rm dis}=7.0\times10^{17}[\rm kg/m^3]$ and $\rho_{\rm inn}^{\rm dis}=9.0\times10^{17}[\rm kg/m^3]$.}
\label{MR_varios_lambda}
\end{center}
\end{figure}

In Fig. \ref{MR_varios_lambda}, the mass-radius curves are compared with the observational data considering $\Gamma_{\rm inn}=\Gamma_{\rm out}=2.4$, some values of $\lambda$ and two $\rho_{\rm inn}^{\rm dis}$. On the top and bottom panel are employed $\rho_{\rm inn}^{\rm dis}=7.0\times10^{17}[\rm kg/m^3]$ and $\rho_{\rm inn}^{\rm dis}=9.0\times10^{17}[\rm kg/m^3]$, respectively. In this figure is used $10^{18}\leq\rho_c\leq10^{20}[\rm kg/m^3]$. The observation data correspond to the NICER constraints obtained from the pulsars PSR J$0030+0451$ \cite{riley2019,miller2019} and PSR J$0740+6620$ \cite{riley2021,miller2021}. The corresponding bands of the pulsars PSR J$0740+6620$ \cite{cromartie2020}, PSR J$0348+0432$ \cite{antoniadis2013} and PSR J$1614+2230$ \cite{demorest2010} are also presented. From the figure, we observe that the change of $\rho_{\rm inn}^{\rm dis}$ affects the stellar structure configuration; being this change more noticeable in the range of low central energy densities. In this interval, for larger $\rho_{\rm inn}^{\rm dis}$, greater total mass and smaller total radius are found. From these results, we note that the change of internal phase transition energy density allows us to obtain some results more accurate and closer to empirical evidence of the neutron stars PSR J$0030+0451$. In addition, we see that with the increment of $\rho_{\rm inn}^{\rm dis}$, grow the possibility of having equilibrium solutions with sharper phase transition (smaller density jump parameter $\lambda$) within PSR J$0030+0451$. On the other hand, from Figs. \ref{r_m} and \ref{MR_varios_lambda}, we observe that in the range of larger total mass, a stiffer fluid in the core could help to reach empirical evidence of neutron stars PSR J$0740+6620$, PSR J$0348+0432$, and PSR J$1614+2230$.

\begin{figure*}[ht]
\begin{center}
\includegraphics[width=0.45\linewidth]{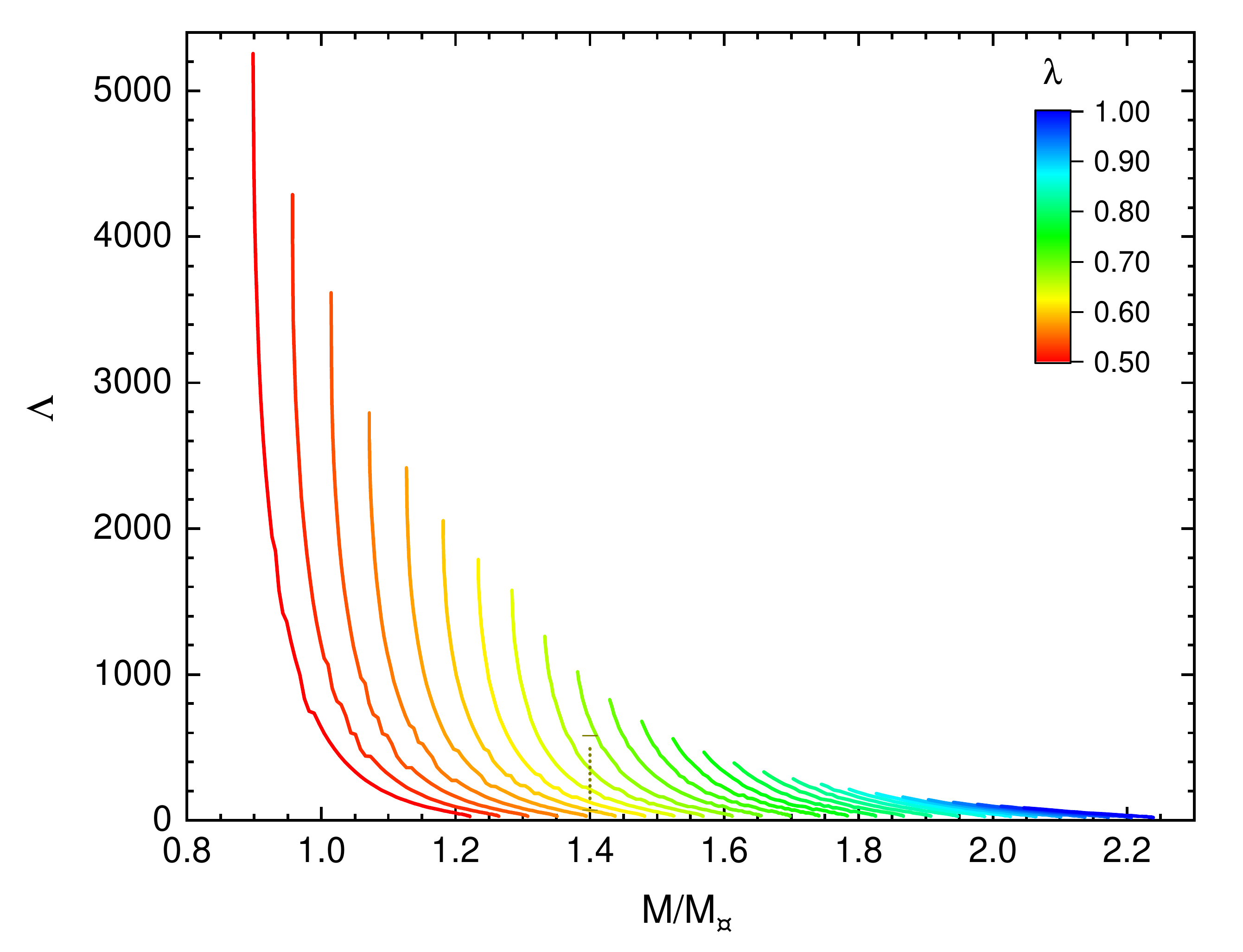}
\includegraphics[width=0.45\linewidth]{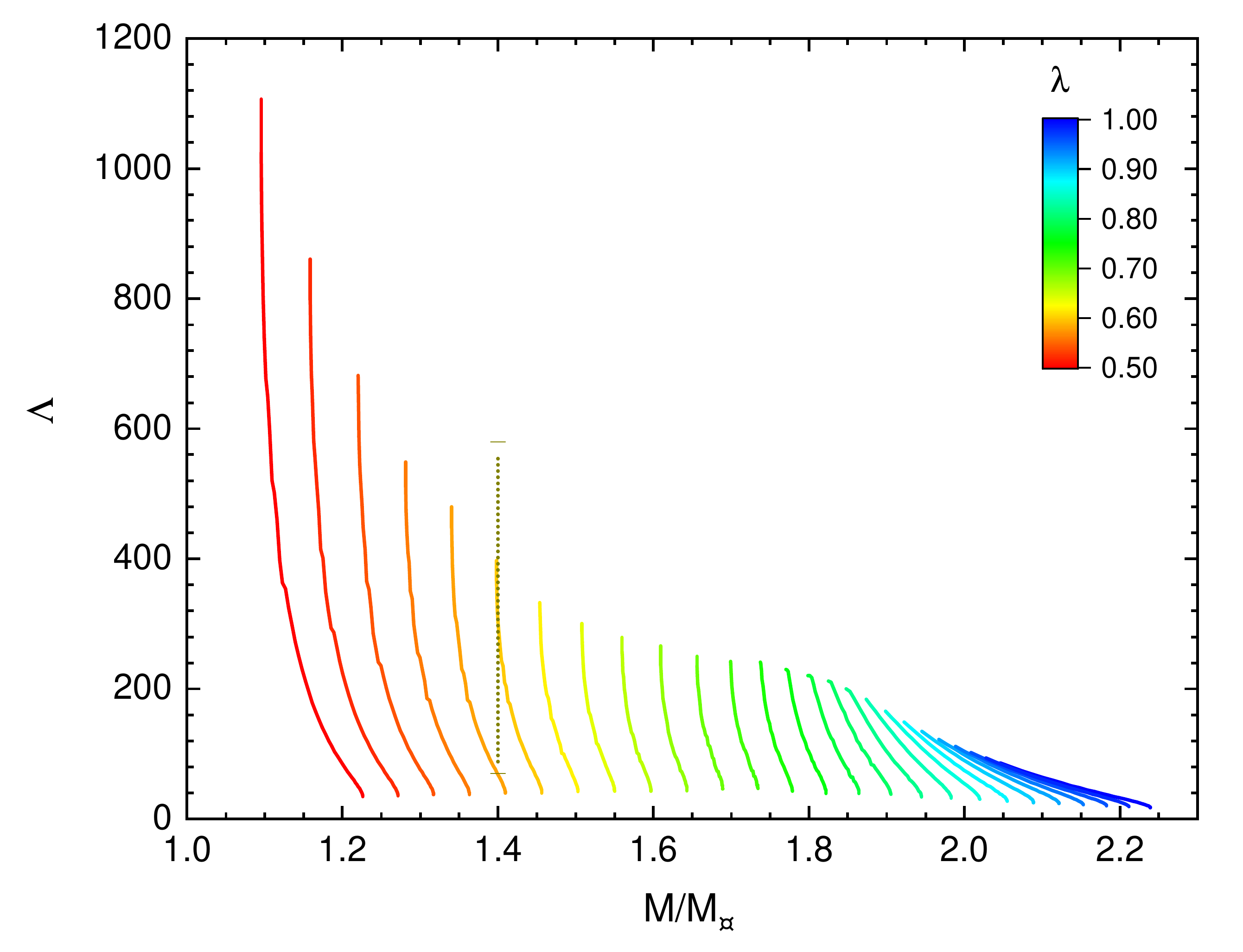}
\includegraphics[width=0.45\linewidth]{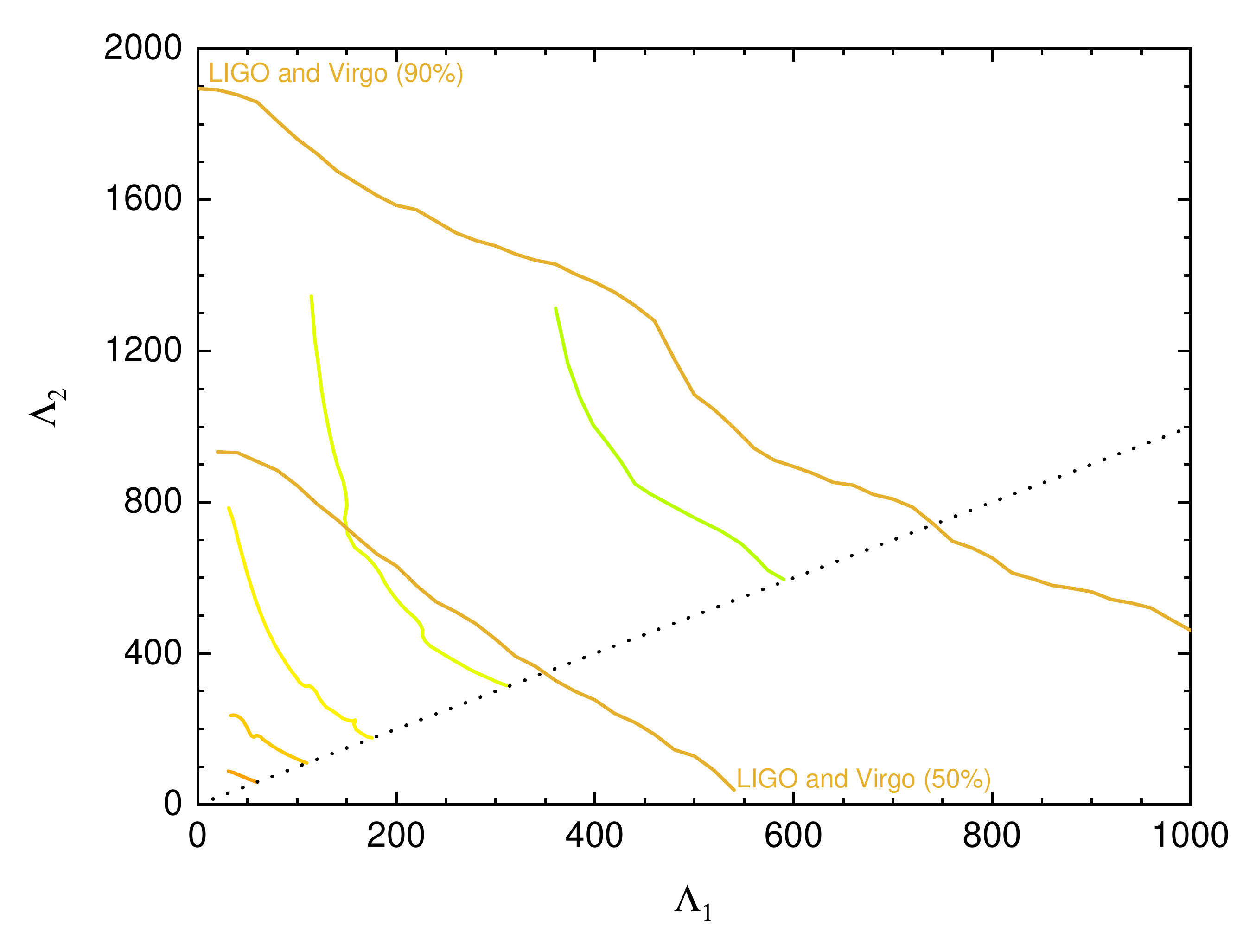}
\includegraphics[width=0.45\linewidth]{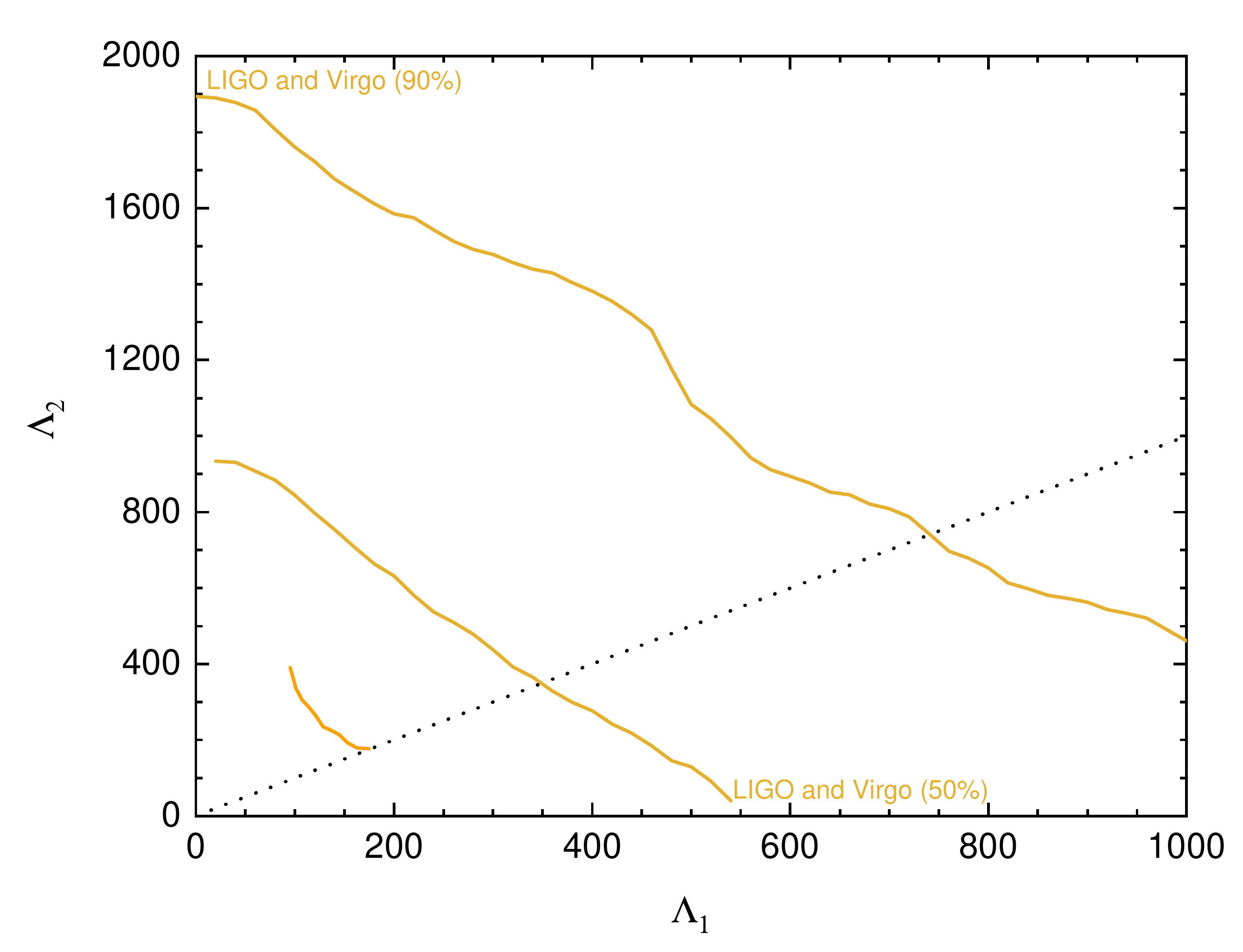}
\caption{Top: Dimensionless tidal deformability $\Lambda$ against the total mass in Sun masses. The vertical line of the dash-dot-dot plots the tidal deformability of the GW$170817$ event measured at \cite{abbott_2018a_tidal}. Bottom: The dimensionless tidal deformability $\Lambda_1$ and $\Lambda_2$ for a binary NS system with masses $m_1$ and $m_2$ and a chirp mass of $1.4M_{\odot}$ considering the combination with $m_1>m_2$. The diagonal short dot line marks $\Lambda_1=\Lambda_2$ limit. The solid top and bottom yellow lines denote respectively $90\%$ and $50\%$ level established by LIGO-Virgo scientific network in the low-spin prior scenario. On the left and right panels are used $\rho_{\rm inn}^{\rm dis}=7.0\times10^{17}[\rm kg/m^3]$ and $\rho_{\rm inn}^{\rm dis}=9.0\times10^{17}[\rm kg/m^3]$, respectively. Only stable equilibrium configurations with rapid conversions at the interface considering $\Gamma_{\rm inn}=\Gamma_{\rm out}=2.4$ are presented.}\label{MLambda_varios_lambda}
\end{center}
\end{figure*}

In Fig. \ref{MLambda_varios_lambda}, the top and bottom panels, respectively, present the tidal deformability against the total mass and $\Lambda_1$-$\Lambda_2$ curves for a binary NS system with chirp mass equal to GW$170817$ considering the relation \eqref{m1m2} where $m_1$ and $m_2$ goes from $1.36M_{\odot}\leq m_1\leq1.60M_{\odot}$ and $1.17M_{\odot}\leq m_2\leq1.36M_{\odot}$. The inner phase transition energy density considered on the left and right panels are $7.0\times10^{17}[\rm kg/m^3]$ and $9.0\times10^{17}[\rm kg/m^3]$, respectively. In all panels of Fig. \ref{MLambda_varios_lambda}, only stable equilibrium configurations for the rapid transition case, employing $\Gamma_{\rm inn}=\Gamma_{\rm out}=2.4$, are considered. As stated above, the observation data correspond to the event GW$170817$.

In Fig. \ref{MLambda_varios_lambda}, on the top panels, it can be seen that all curves tidal deformability ($\Lambda$)-total mass ($M/M_{\odot}$) decay with the increment of $\rho_{\rm inn}^{\rm dis}$. From this, we understand that this phenomenon allows us to have equilibrium configurations with a lower density jump parameter $\lambda$ (a sharper phase transition) within the observational data of the GW$170817$ event. From the bottom panels, we note that some equilibrium solutions located outside the range of observational data derived from the GW$170817$ event fall within this interval when we increase the phase transition energy density $\rho_{\rm inn}^{\rm dis}$. In addition, from Figs. \ref{MLambda_varios_lambda} and \ref{TD_masa}, we can also say that having a stiffer fluid in neutron stars' cores helps us to have static equilibrium configurations within the range of observational data.

\section{Conclusions}\label{conclusion}

In this work, we investigated the influence of the phase transition on the equilibrium, radial stability, and tidal deformability of NSs with a stiffer fluid in the core. In the core and the envelope of the star, the relativistic polytropic equation of state is considered. The spherical equilibrium configurations are connected smoothly with the Schwarzschild exterior spacetime. We examined the change of the mass, radius, speed of sound, core radius, the eigenfrequency of the fundamental mode of the star with a slow and rapid phase conversion at the interface, and tidal deformability for different density jump parameters $\lambda$, phase transition energy densities $\rho_{\rm inn}^{\rm dis}$, interior polytropic exponents $\Gamma_{\rm inn}$, and exterior polytropic exponent $\Gamma_{\rm out}=2.4$.

As well as in the study of NSs developed in \cite{sotani_2001}, which employs a non-relativistic polytropic equation of state, we note that some aspects of the static equilibrium configurations -such as the mass and radius- are affected by the phase transition, stiffer fluid in the core (which change with $\Gamma_{\rm inn}$), and phase transition energy density. 

For the values $\Gamma_{\rm inn}$ and $\lambda$ employed, in the slow case, the zero eigenfrequencies of the fundamental mode are attained beyond the maximum mass points and, in the rapid case, the maximum masses points mark the beginning of the radial instability thus indicating that the regions constituted by stable and unstable stars can be recognized by the conditions $dM/d\rho_c>0$ and $dM/d\rho_c<0$, respectively. These results are in concordance with those one reported in the works \cite{pereira_flores2018,tonetto_2020,mariani_lugones2019}. 

The change of the tidal deformability for a NS ($\Lambda$) and a binary NS system ($\Lambda_1$ and $\Lambda_2$), with equal chirp mass as GW$170817$ event, as a function of $\Gamma_{\rm inn}$ and $\lambda$, has been analyzed. We obtained a dependence of the dimensionless tidal deformability with these two factors in the aforementioned frameworks. For NS configurations, for some interval of masses, we noted that $\Lambda$ grows and decreases with the increment of $\Gamma_{\rm inn}$ and diminution of $\lambda$. In turn, for a binary NS scenario, we showed that the phase transition and stiffer fluid in the NS core could also play an important role in the detection of NSs. These results are in agreement with those ones published in \cite{parisi_2021}.

We also investigated the dependence of some physical parameters of NSs with the phase transition energy density. At the interval of low central energy densities, we found that $\rho_{\rm inn}^{\rm dis}$ can also be important in the study of NSs since their physical parameters could be significantly affected by the value of phase transition energy density. 

Finally, we noted that a change in jump density parameter $\lambda$, phase transition energy density $\rho_{\rm inn}^{\rm dis}$, and stiffer core fluid $\Gamma_{\rm inn}$ could lead to the possibility that some equations of state that are outside of the observational data, can be inside this framework for accurate values of $\lambda$, $\rho_{\rm inn}^{\rm dis}$ and $\Gamma_{\rm inn}$.

\begin{acknowledgements}
JDVA would like to thank the Universidad Privada del Norte and Universidad Nacional Mayor de San Marcos for funding - RR Nº$\,005753$-$2021$-R$/$UNMSM under project number B$211\-31781$. CHL is thankful to the Fundação de Amparo à Pesquisa do Estado de São Paulo (FAPESP) under thematic project $2017/05660-0$, Grant No. $2020/05238-9$ and Power Data Tecnologia Ltda for providing a technological environment for data processing.

\end{acknowledgements}

\end{document}